%% file: arxiv.tex
\documentclass[preprint]{imsart}

\RequirePackage{amsthm,amsmath,amsfonts,amssymb}
\RequirePackage[numbers]{natbib}
\RequirePackage[colorlinks,citecolor=blue,urlcolor=blue]{hyperref}%
\RequirePackage{graphicx}%

\RequirePackage[capitalize]{cleveref}

\startlocaldefs
\RequirePackage[margin=1in]{geometry}

\usepackage{tikz}
\usepackage{overpic}
\usetikzlibrary{positioning}

\newcommand{\rev}[1]{\textcolor{black}{#1}}
\def\ps@imsheadings{%
 \def\@oddfoot{\hfill\info@line}\let\@evenfoot\@oddfoot
 \def\@oddhead{\runninghead@size\hfill\rightmark\hfill\llap{\pagenumber@size\thepage}}%
 \let\@evenhead\@oddhead}
\endlocaldefs

\input{macros}

\begin{document}

\begin{frontmatter}
\title{Transport for Conditional Sampling:\\ Foundations and Applications}
\runtitle{Transport for Conditional Sampling}

\begin{aug}
\author[A]{\fnms{Ricardo}~\snm{Baptista}\ead[label=e1]{r.baptista@utoronto.ca}}
\author[B]{\fnms{Benjamin}~\snm{Bloem-Reddy}\ead[label=e2]{benbr@stat.ubc.ca}}
\author[C]{\fnms{Bamdad}~\snm{Hosseini}\ead[label=e3]{bamdadh@uw.edu}}
\address[A]{Department of Statistical Sciences, University of Toronto\printead[presep={\ }]{e1}.}
\address[B]{Department of Statistics, University of British Columbia\printead[presep={\ }]{e2}.}
\address[C]{Department of Applied Mathematics, University of Washington\printead[presep={\ }]{e3}.}
\end{aug}

\begin{abstract}
\input{abstract}

\end{abstract}


\end{frontmatter}

\input{main_text}

\bibliographystyle{imsart-number} %
\bibliography{references}       %

\end{document}

%% file: macros.tex
\newcommand{\x}{x}
\newcommand{\y}{y}
\newcommand{\z}{z}
\renewcommand{\u}{u}
\renewcommand{\v}{v}
\newcommand{\w}{w}
\newcommand{\X}{X}
\newcommand{\Y}{Y}
\newcommand{\Z}{Z}
\newcommand{\U}{U}
\newcommand{\V}{V}
\newcommand{\W}{W}

\newcommand{\Xc}{\mathcal{X}}
\newcommand{\Yc}{\mathcal{Y}}
\newcommand{\Zc}{\mathcal{Z}}
\newcommand{\Wc}{\mathcal{W}}

\newcommand{\D}{\mathsf{D}}
\newcommand{\dkl}{\D_{\textrm{KL}}}
\newcommand{\KR}{\textrm{KR}}
\newcommand{\MMD}{\textrm{MMD}}
\newcommand{\BT}{\textrm{BT}}
\newcommand{\BKR}{\textrm{BKR}}

\newcommand{\R}{\mathbb{R}}
\renewcommand{\P}{\mathbb{P}}

\newcommand{\dd}{{\rm d}}

\newcommand{\hatT}{\widehat{T}}

\newcommand{\source}{\rho}
\newcommand{\target}{\mu}

\usepackage{mathtools}

\newtheorem{theorem}{Theorem}
\newtheorem{corollary}{Corollary}

\newtheorem{example}{Example}
\newtheorem{remark}{Remark}

\newcommand{\pa}{\mathrm{pa}}
\newcommand{\noise}{w}
\newcommand{\Noise}{W}

\newcommand{\bfN}{\Noise}
\newcommand{\bff}{T}
\newcommand{\bfX}{Z}
\newcommand{\bfx}{z}

\DeclareMathOperator*{\argmin}{arg\,min}

%% file: abstract.tex
\rev{Conditional sampling is central to Bayesian inference, generative modeling, and causal inference: given a joint law for random variables $\X$ and $\Y$ and an observation $\Y=\y$, one seeks to sample from the conditional law of $\X\mid\Y=\y$. Even when the joint distribution is accurately represented, this task can be computationally difficult as it may require evaluating intractable normalizing constants or solving a new inference problem for each observation. Measure transport provides an approach to this problem by identifying a map with triangular structure that directly pushes forward a reference to the conditional of a joint distribution for any observation. This article reviews the theoretical, methodological, and algorithmic foundations of triangular transports, with the Knothe–Rosenblatt rearrangement as a canonical construction. We discuss methods for estimating triangular maps from finite samples, as well as their use in simulation-based inference and causal inference. While our primary aim is to review existing formulations, we also present new ideas and open questions for conditional sampling using transport.}

%% file: main_text.tex
\section{Introduction}\label{sec:intro}

Conditional probability distributions play a central role in Bayesian inference, supervised learning, generative modeling and causal inference. Characterizing conditional distributions enables several statistical tasks including estimating parameter uncertainty from data and forecasting states of a system.  
Although such problems have a long history in statistics and related areas, in recent years new approaches based on measure transport have emerged as important alternatives to traditional conditional sampling algorithms. The aim of this article is to review recent progress in this rapidly changing area and in particular to highlight the mathematical, computational and statistical properties of \emph{triangular} transport maps that have been largely 
developed within the applied mathematics and machine learning literature for this problem.

To set the basic notation, let $(\Y, \X)$ be random variables \rev{taking values in spaces $\Yc$ and $\Xc$} and distributed according to the joint distribution $\target_{\Y,\X}$. 
We denote the conditional distribution of $\X$ given $\Y$ under the joint law by $\target_{\X|\Y}$.
For simplicity, we assume that $\target_{\X|\Y}$ admits a density with respect to the Lebesgue measure, which we denote, with a slight abuse of notation, by 
$$\target_{\X|\y}(\x) \coloneqq \target_{\X|\Y}(\x \mid \y),$$
for each fixed $\y \in \Yc$. We will understand these conditional probability distributions 
in the sense of {\it regular conditional distributions} (i.e., Markov transition kernels) 
as they are often studied in statistics and probability \cite[Ch.~10]{bogachev2007measure-II}.
\rev{Throughout this article, we  mainly consider the 
setting where $\Yc, \Xc$ are finite-dimensional Euclidean spaces unless stated otherwise. We note that the extension of the ideas and results discussed in the paper to infinite-dimensional vector spaces is possible, but largely open, as we will briefly discuss in Section~\ref{sec:extended_methodology}.}

Measure transport provides a framework for characterizing a complex conditional distribution 
as the transformation of a tractable reference distribution $\source$ that is chosen to be easy to sample, e.g. a standard Gaussian. These transformations are the topic of this 
review and are the main tools that we try to understand and develop. The basic idea of transport 
is to construct a family of \emph{transport maps}  
$T_\y\colon \Xc \to \Xc$ 
(viewing $\y$ as a parameter) such that $T_\y \# \source = \target_{\X \mid y}$, 
where $\#$ is 
{\it the pushforward 
operator}: 
\begin{equation}\label{parametric-conditional-maps}
    T_\y \# \source = \text{Law} \{ T_\y(\U) \mid \U \sim \source \}.
\end{equation}
The convenience of identifying $T_\y$ is that we can generate i.i.d.\thinspace samples from the conditional $\mu_{\X \mid \y}$ simply by drawing samples $\u^{(i)} \sim \source$ and evaluating $T_y(\u^{(i)})$. 

Next,~\Cref{sec:motivation} outlines various domains where conditional distributions need to be identified and where transport maps have wide potential applications. \Cref{sec:triangular_transport} introduces the class of triangular transport maps, which are well-suited for conditional sampling. Section~\ref{sec:notation} presents the outline for the rest of the article.

\subsection{Motivating Applications} \label{sec:motivation}

\subsubsection{Bayesian Inverse Problems} \label{sec:BIP}

Conditional distributions naturally arise in the context of the Bayesian approach to solving inverse problems~\cite{stuart2010inverse,sanz2023inverse,bingham2024bip}. Here, $\y$ are typically noisy and indirect observations associated with 
an unknown parameter $\x$.\footnote{The common statistics notation for the parameter is $\theta$; we use $\x$ for consistent notation with other applications.} For example, we may have the model $\y = G(\x) + \eta$, where $G$ is a forward model such as a numerical solver for an ordinary or partial differential equation or a simulator 
and $\eta$ represents additive measurement noise with a known distribution \rev{with $\eta \perp \X$}. The goal of Bayesian inference is 
to characterize the conditional distribution of $\x$ given an instance $\y$ while modeling $\x$ as a random variable $\X$ that is a-priori distributed 
according to its {\it prior distribution} $\target_\X$. Under this 
assumption, we can then define the joint distribution
$$
\target_{\Y, \X}:= \text{Law} 
\{ (Y,X) \mid X \sim \target_\X, \:\: \Y = G(X) + \eta \}.
$$
The conditional distribution of interest in this problem is $\target_{\X \mid y}$, the {\it posterior distribution}.

For example, when $\eta$ is a centered Gaussian variable with covariance $\Gamma$, it is easily verified that the likelihood takes the form $Y \mid X=x \sim \mu_{\Y|\X}(\cdot|x) = \mathcal{N}(\cdot;G(x), \Gamma)$.  
Then, Bayes' rule gives the familiar expression for the density of the posterior distribution
$$
\target_{\X \mid \Y}(\x \mid \y) \propto \exp\bigg(-\frac{1}{2}\|y - G(x)\|_{\Gamma}^2\bigg) \target_\X(\x) \;.
$$
For non-Gaussian priors $\target_\X$ and/or nonlinear forward models $G$, sampling from the posterior distribution is intractable without iterative methods such as Markov chain Monte Carlo (MCMC)~\cite{robert1999monte}. While MCMC is attractive theoretically, its practical utility can be limited in these situations. Each MCMC iteration typically requires one or more evaluations of the likelihood, which in this case involves a call to the forward model $G$. This can be prohibitive, especially when MCMC mixes slowly. 

\rev{Beyond its computational cost,} MCMC is challenging to implement when the likelihood $\target_{\Y|\X}$ is expensive to evaluate or even intractable to compute. 
This may occur when the forward model $G$ depends on both the parameter $x$ and a \emph{latent variable} $z$ (also referred to as a nuisance variable) with distribution $\mu_{\Z|\X}(\cdot|x)$. Here, likelihood function evaluations require computing the (possibly high-dimensional) integral
$$\mu_{\Y \mid \X}(\y|\x) = \int \mu_{\Y \mid \Z,\X}(\y\mid \z, \x) \mu_{\Z\mid\X}(\z|\x)d\z.$$
In this setting, likelihood-free inference~\cite{cranmer2020frontier, sisson2018handbook} is an alternative to MCMC that avoids evaluating this likelihood function. These approaches characterize the 
posterior distribution $\mu_{\X|\Y}$ given \emph{only} samples of the joint distribution $\mu_{\Y,\X}$; see 
Section~\ref{sec:conditional_modeling} for more details. 

\subsubsection{Sequential Inference} \label{sec:sequential_inference}

A more complex inverse problem is to infer a sequence of time-dependent states $(\X_t)$ of a dynamical system from observations $(\Y_t)$, often collected sequentially in time. This problem lies at the core of state estimation in many application domains 
including robotics \cite{thrun2002probabilistic}, 
finance~\cite{Triantafyllopoulos_2021}, 
ecology~\cite{augermethe2021ssm_ecology} and 
geophysics~\cite{calvello2022ensemble}. One popular state estimation problem is \rev{discrete-time} filtering, which recursively approximates a conditional distribution for the states given observations from the Markovian state-space model
\begin{align*}
    \X_t &= \Psi(\X_{t-1}) + \eta, \quad X_0 = x_0 \\
    \Y_t &= H(\X_t) + \epsilon,
\end{align*}
where $\Psi$ and $H$ are dynamics and observation operators, respectively, $\eta$ and $\epsilon$ represent independent noise variables in the dynamics and measurements, and $x_0$ is the (possibly random) initial condition. Under this model, the conditional distribution of interest is the {\it filtering distribution}
$\mu_{\X_t|\y_{1:t}}$ for the state $\X_t$ given the sequence of past observations $\y_{1:t}$. The filter is known to satisfy a two-stage recursive formula to update $\mu_{\X_{t}|\y_{1:t}}$ to $\mu_{\X_{t+1}|\y_{1:t+1}}$ by predicting the state and applying Bayes' rule:
\begin{align*}
    \mu_{\X_{t+1} \mid \y_{1:t}}(\x_{t+1}) &= \int \mu_{\X_{t+1} \mid \X_{t}}(\x_{t+1} \mid \x_{t}) \mu_{\X_{t}|\y_{1:t}}(\x_{t})d\x_{t}\\
    \mu_{\X_{t+1} \mid \y_{1:t+1}}(\x_{t+1}) &\propto \mu_{\Y|\X}(\y_{t+1} \mid \x_{t+1})\mu_{\X_{t+1} \mid \y_{1:t}}(\x_{t+1}).
\end{align*}

This decomposition reduces the filtering 
problem to solving a sequence of 
inverse problems at each step, with $\mu_{X_{t+1} \mid \y_{1:t}}$
viewed as the prior and $\mu_{X_{t+1} \mid \y_{1:t+1}}$ 
as the posterior. As in~\Cref{sec:BIP}, when the dynamic and observation operators are non-linear, the filtering distributions are generally non-Gaussian. 
One additional challenge in filtering is that the prior distribution (arising from the filter in the previous step) is typically only available via a collection of samples. This restricts the use of MCMC or variational inference methods that require access to the prior density. As a result, sequential inference is often performed by instead finding a transport to directly map prior samples into posterior samples at each step; see Section~\ref{sec:conditional_modeling}.

\subsubsection{Causal Modeling and Inference}

The primary objective of causal modeling is to estimate quantities from a distribution that differs from an observed distribution, and where the difference is the result of an intervention. For example, one might estimate the distribution of health outcomes under dosage levels or treatment policies that are different from those observed. If the  distribution of interest corresponds to \rev{new \emph{units} (e.g., patients) 
that are independent of those in the observed sample then we say that it is an \emph{intervention distribution}}. If, on the other hand, the distribution of interest corresponds to the \emph{same} units under the intervention then it is called a \emph{counterfactual distribution}.\footnote{In the potential outcomes framework, the interventional distribution is often called the counterfactual distribution. We use the language of \cite{Pearl_2009} for ease of presentation in \cref{sec:causal_inference}. The difference is whether or not the observational and interventional distributions are coupled.} The core statistical question is to estimate shifts of conditional distributions. 

To be more precise, consider $\X$ to be the variable of interest 
that is caused by $\Y$, the variable 
that can be modified due to an intervention or counterfactual study. Then our goal is to 
model changes in $\mu_{\X \mid \y}$ due to changes in the 
value of $\y$. In general, we would not only consider 
a pair of $(\X, \Y)$ but rather collections of these variables with 
their causal relations encoded in a graph; see \Cref{sec:causal_inference}.

\subsection{Triangular Measure Transport} \label{sec:triangular_transport}

A common feature of the examples introduced in Section~\ref{sec:motivation}, among others, is that accessing samples from the conditional distributions $\mu_{\X \mid \y}$ for different values of $\y$ is computationally prohibitive. This makes it challenging to identify the entire family of transport maps $T_\y$ that can be used to sample from an arbitrary conditional distribution for any new value $\y$.
In~\Cref{sec:transport_map} we will show how to identify such a family of
transports using only samples from the joint distribution
$\target_{\Y, \X}$ instead, which are often accessible. This will be achieved by considering a family of \emph{triangular transport maps}, whose specific structure for the dependence on $\y$ enables the direct sampling from the conditionals of a joint distribution as in~\eqref{parametric-conditional-maps}. Interestingly, this 
result will allow us to solve many of the motivating 
problems outlined in the previous section. 

The goal of this review is to provide a concise overview on the foundations of triangular transport maps and their applications. We will discuss their construction and mathematical properties in addition to demonstrating their use
for inference 
and causal modeling. 
\rev{We also highlight two related review articles~\cite{marzouk2016sampling, ramgramber2025} as complementary references. Both articles focus on the construction of (strictly triangular) Knothe--Rosenblatt 
maps on finite-dimensional Euclidean spaces with applications to Bayesian inference. The present review differs in three respects: (i) we present a generalization to block-triangular maps (\Cref{sec:transport_map}); 
(ii) we focus on the approximation-theoretic and statistical estimation properties of triangular maps (\Cref{sec:computation}); and (iii) we provide connections between triangular transport and causal modeling (\Cref{sec:causal_inference}) in addition to Bayesian inference (\Cref{sec:conditional_modeling}).} 

\subsection{Organization and Notation} 
\label{sec:notation}
The remainder of this article is organized as follows. \Cref{sec:transport_map} introduces triangular transport maps and their analytical properties. \Cref{sec:computation} shows how to learn triangular maps from data, including their statistical estimation properties. Sections~\ref{sec:conditional_modeling} and \ref{sec:causal_inference} describe applications of the methodology to Bayesian inference and causal modeling, respectively. \rev{\Cref{sec:extended_methodology} presents an extension of triangular maps to more general, potentially infinite-dimensional, spaces.} In~\Cref{sec:outlook} we introduce open questions and research directions. References to related works in the literature will appear in each relevant section.

Throughout the article we mostly assume that $\Xc, \Yc$
are finite-dimensional Euclidean spaces $\R^d$ and $\R^m$, respectively, and that measures
on these spaces are absolutely continuous with respect to the Lebesgue measure. 
For a transport map defined on $\Xc$, we will say that $T$ pushes forward $\source$ to $\target$ if $T(X) \sim \target$ for $X \sim \source$. This means that $\target(A) = \source(T^{-1}(A))$ for all Borel sets $A \subset \Xc$, which we denote as $T \# \source = \target$. 
We often need to distinguish the components of the random variables 
distributed according to the reference distribution.  
In such scenarios, we use the notation $(\V, \U)$ to denote reference random variables that will be pushed forward to target random variables $(\Y, \X)$. We will also use the lowercase letters $(\v, \u)$ and, respectively, $(\y, \x)$ to denote points in the domain and range of transport maps to emphasize their connections to the subsequent random variables. Occasionally, we also use 
$\W = (\V, \U) \in \Zc :=( \Yc \times \Xc)$ to denote the concatenation of 
$\V, \U$ and similarly $\Z = (\Y, \X) \in \Zc$. 

\section{Conditional Transport Maps} \label{sec:transport_map}

In this section we formally introduce triangular transport maps for characterizing conditional distributions. We start by recalling transport maps in one-dimension and extend them to one-dimensional conditional distributions in \Cref{sec:one_dimensional_map}. We use this idea to build a single triangular map from a collection of these maps in 
\Cref{sec:KR_map}. This triangular map is also known as the Knothe--Rosenblatt rearrangement. We then extend this idea to block-triangular maps for multivariate random variables $(\Y,\X)$ in \Cref{sec:block-maps}. Finally, we discuss
some properties of triangular maps, 
such as their regularity, in~\Cref{sec:regularity}. 

\subsection{One-Dimensional Transport Maps} \label{sec:one_dimensional_map}

We first recall the monotone transport map between one-dimensional distributions 
$\source_\U,\target_\X \in \mathcal{P}(\R)$. Let $F_{\source}$ and $F_{\target}$ denote the cumulative distribution functions (CDF) of $\source_\U$ and $\target_\X$, respectively, i.e., $$F_\source(u) = \int_{-\infty}^u \source_\U(t)\dd t.$$ For an atomless distribution $\source_\U$, the CDF defines a continuous 
transport map pushing forward $\source_\U$ to the uniform distribution $\mathcal{U}[0,1]$. Moreover, the inverse CDF, or quantile function,  $F_\source^{-1}$ pushes forward $\mathcal{U}[0,1]$ to $\source_\U$. Repeating this for $\target_X$ allows us to define the transport map 
\begin{equation} \label{eq:OneDimMap}
T(u) = F_\target^{-1} \circ F_\source(u),
\end{equation}
that pushes $\source_\U$ to $\target_\X$; 
see \cite[Theorem~2.5]{santambrogio2015optimal} for a proof.

The map defined in~\eqref{eq:OneDimMap} is 
amenable to computation. In particular, if $\source_\U$ and $\target_\X$ are specified using samples, then an estimator of $T$ is given by replacing the CDFs by their empirical estimators~\cite{bobkov2019one}. This map has other 
interesting properties, notably, it is optimal (in the sense of optimal 
transport theory) for any strictly convex cost function in 1D~\cite{santambrogio2015optimal} and it is
the \rev{unique monotone increasing} transport from $\source_\U$
to $\target_\X$.

Let us now consider one-dimensional transport between conditional distributions 
$\source_{\U \mid y}, \target_{\X \mid y} \in \mathcal{P}(\R)$ for different 
values of $y$. Let $F_\source(\cdot \mid y), F_\target(\cdot \mid y)$ denote the CDFs of these conditionals. Under the same atomless condition on $\source_{\U \mid y}$, the composition 
\begin{equation} \label{eq:composed_map}
T_{\y}(\u) := F_\target(\cdot \mid \y)^{-1} \circ F_\source(\u \mid \y)
\end{equation}
defines a parametric family of monotone maps 
pushing forward $\source_{\U \mid \y}$ to $\target_{\X \mid \y}$ \rev{for each $\y$}. These 
maps are precisely the one-dimensional versions of the maps that we introduced in~\eqref{parametric-conditional-maps}.
We note that in many practical applications, such as generative modeling or Bayesian inverse problems, we have freedom to choose the reference distribution $\source_{\V, \U}$ and certain choices can simplify the maps $T_\y$. Most notably, we can take 
$\source_{\V, \U} =  \source_\V \otimes \source_\U$, i.e., $\V, \U$ are independent under the reference distribution, in which 
case $F_\source(\u \mid \y) = F_\source(\u)$ and  $T_\y(\u) = F_\target^{-1}(\cdot \mid  \y) \circ F_\source(\u)$ is the transport map pushing the $\U$ marginal $\rho_\U$ to 
the target conditional $\target_{\X \mid y}$.
We now outline an example where the maps $T_\y$ can be characterized analytically, and an example to visualize families of conditional maps.

\begin{example}[Location-Scale Model]
\label{ex:loc:scale}
Let $\X \sim \pi$ and ${m(\y) + b(\y) \cdot \X}$ be a collection of random variables from a location-scale family with location parameter $m(y)$ and scale parameter $b(y) > 0$ indexed by $\y$. If $\source_{\U|\y}$ and $\mu_{\X|\y}$ are members of this class with parameters $m_{\source}, b_{\source}$ and $m_{\mu}, b_{\mu}$, respectively, then their CDFs take the form
\begin{align*}
F_\source(\u \mid y) &= F_\pi((\u - m_\source(y))/b_{\source}(y)) \;, 
\end{align*}
and similarly for $F_{\mu}$, 
where $F_\pi$ is the CDF of $\pi$. Using the composition in~\eqref{eq:composed_map}, the transport map
\begin{align*}
    T_\y(\u) = m_\target(y) + \frac{b_{\mu}(y)}{b_{\source}(y)}(\u - m_\source(y)) \;
\end{align*}
pushes $\source_{\U|\y}$ to $\mu_{\X|\y}$ for each $\y$. This map is affine in $\u$ and transports between any two members of the same family.
\end{example}

\begin{example}[Multimodal Conditional] Let the target distribution $\mu_{\Y,\X}$ be specified by a Gaussian marginal $\X \sim \mathcal{N}(0,0.25)$ and the quadratic forward model $\Y = \X^2 + \eta$ for $\eta \sim \mathcal{N}(0,0.01)$ as in~\Cref{sec:BIP}. The first and second rows of Figure~\ref{fig:example_cde} plot the contours for the joint density of $\mu_{\Y,\X}$ and the conditional density of $\mu_{\X|\y}$ for conditioning values $\y \in \{0,0.25,0.5\}$. The conditional distribution smoothly varies between being unimodal for $\y = 0$ and bimodal for $\y \in \{0.25,0.5\}.$ The third row of Figure~\ref{fig:example_cde} displays the corresponding three transport maps $T_\y$ that push forward a Gaussian  reference $\source_\U = \mathcal{N}(0,1)$ to each conditional $\mu_{\X|\y}$. The family of maps inherits the smooth dependence on the conditioning value $\y$. 
\vspace{-0.5cm}
\input{diagram}
\end{example}

\subsection{The Knothe--Rosenblatt Map} \label{sec:KR_map}

The one-dimensional maps outlined in the previous section can be extended to define transport maps between multivariate probability distributions while maintaining their conditioning properties. Since we will be working with multi-variate random variables in this section we will use $\Z$ and $\W$
to denote the target and reference random variables, respectively, following our convention in \Cref{sec:notation}.

The law of total probability~\cite[Section~8]{shiryaev2016probability} (i.e., disintegration)  allows us to represent any distribution $\target_\Z \in \mathcal{P}(\R^d)$ in terms of its marginal conditionals. More precisely, for 
$\Z = (\Z_1, \dots, \Z_d) \sim \target_\Z$, we have 
the decomposition
\begin{equation}\label{total-probability-R-d}
\target_\Z (\z) = \target_{\Z_1}(\z_1)
\target_{\Z_2 \mid \z_1}(\z_2)
\cdots \mu_{\Z_d \mid \z_{1:d-1}}(\z_d),
\end{equation}
where we used the shorthand notation $\z_{1:k} 
= (\z_1, \dots, \z_k) \in \R^k$.
For simplicity, we consider an i.i.d. 
reference distribution of the form 
$\source_\W(\w) = \bigotimes_{i=1}^d \varrho(\w_i)$ for some one-dimensional 
atomless distribution $\varrho \in \mathcal{P}(\R)$, such as a standard normal. Then we can explicitly construct a transport map $T_{\KR}:\R^d \to \R^d$ 
of the form 
\begin{equation} \label{eq:KR}
  T_{\KR}(\w) =  
  \bigl(T_1(\w_1), T_2(\z_1, \w_2), \dots, T_d(\z_{1:d-1}, \w_d) \bigr)
\end{equation}
with components 
\begin{equation} \label{eq:KRcomponent_parentdependence}
    T_k(\z_{1:k-1}, \w_k) := F_\target(\cdot \mid \z_{1:k-1})^{-1} \circ F_\source(\w_k),
\end{equation}
where we used 
$F_\target(\cdot \mid \z_{1:k-1})$
to denote the CDF of $\target_{\Z_k \mid \z_{1:k-1}}$ and the conditioning variable $\z_{1:k-1}$ is the output of the previous
components of the map itself. In practice, unrolling this recursion lets us
write the map directly as a function from $\w$ to $\z$, namely $\z = T_{\KR}(\w)$, where the $k$-th component depends only on $\w_{1:k}$. This is the form we work with in the remainder of this section.

Observe that the components $T_k$ defined above are nothing but the one-dimensional maps that transport $\varrho$ to each conditional 
$\target_{\Z_k \mid \z_{1:k-1}}$ in~\eqref{total-probability-R-d}, 
and so it is easily verified that 
$T_{\KR} \# \source_\W = \target_\Z$.
The map defined in~\eqref{eq:KR} is known as the 
Knothe--Rosenblatt (KR) map~\cite{rosenblatt1952remarks,knothe1957contributions}. We 
refer to this map as a (fully) triangular map 
since its Jacobian $\nabla T_{\KR}(\w) \in \R^{d \times d}$ is a triangular matrix. 

The construction of the KR map is not restricted to 
the case where $\source_\W$ is i.i.d. as assumed above. In fact, 
if we drop this assumption then the KR map can still 
be defined as in \eqref{eq:KR}, except that the components 
$T_k$ now take the form 
\begin{equation}\label{KR-map-component}
   T_k(\w_{1:k}) = 
   F_\target( \cdot \mid T_{1:k-1}(\w_{1:k-1}))^{-1} 
   \circ F_\source(\w_k \mid \w_{1:k-1}),
\end{equation}
In this case, the $T_k$ are one-dimensional transport maps 
pushing forward $\source_{\W_k \mid \w_{1:k-1}}$ 
to $\target_{\Z_k \mid T_{1:k-1}(\w_{1:k-1})}$ where we use the shorthand $T_{1:k-1}$ to denote the first $k-1$ map components; 
see also~\cite[Section~2.3]{santambrogio2015optimal} for more 
details and discussion about this extension.
%
%
We now provide an example where the KR map has an explicit analytic form.

\begin{example}[Affine triangular maps] \label{ex:affine_maps}
Let $\target_\Z$ be a multivariate Gaussian target $\mathcal{N}(m_\target,\Sigma_\target)$ with mean $m_\target \in \R^d$ and positive definite covariance matrix $\Sigma_{\target} \in \R^{d \times d}$. It is well known that we can generate samples from $\target_\Z$ by computing the Cholesky factorization 
$\Sigma_\target = L_\target L_\target^\top$ and applying the transformation
\begin{equation} \label{eq:GaussianTransformation}
L_\target \W + m_\target \sim \mu_{\Z}, \quad \W \sim \mathcal{N}(0,I_d).
\end{equation}
The lower-triangular Cholesky factor  defines a triangular transport map pushing forward the standard Gaussian reference $\source_\W$ to the target $\target_\Z$. 

When the reference $\source_\W$ is a non-standard Gaussian 
$\source_\W = \mathcal{N}(m_\source,\Sigma_\source)$ where the covariance has the Cholesky factorization $\Sigma_\source = L_\source L_\source^\top$, then the map 
\begin{align*} 
    T_{\KR}(\w) = m_\target + L_{\target}L_\source^{-1}(\w - m_\source) \;,
\end{align*}
pushes forward $\source_\W$ to $\target_\Z.$ This map is given by composing the inverse of transformation~\eqref{eq:GaussianTransformation}, which maps $\source_\W$ to a standard Gaussian, with the transformation~\eqref{eq:GaussianTransformation} that maps a standard Gaussian to $\target_\Z$. Since $L_{\target}L_\source^{-1}$ is lower triangular, this map is triangular and hence it is the KR map.
\end{example}

\Cref{ex:affine_maps} is a generalization of the univariate location-scale target in~\Cref{ex:loc:scale} to the multivariate setting. More broadly, one can show that an affine KR map can be used to map between any members of the same multivariate location-scale family. This idea also applies to group families that contain a triangular subgroup; see \cite{dance2024causalinferencecocycles}.

KR-maps offer several practical advantages. First, the inverse $T_{\KR}^{-1}$ of a bijective KR map can be easily computed by solving a sequence of one-dimensional root-finding problems. 
To see this observe that we can solve $\z = T_{\KR}(\w)$ by solving the one-dimensional problems 
$\z_k  =T_k(\w_{1:k-1},\w_k)$ for $w_k$ sequentially for $k = 1,\dots,d$, 
akin to forward substitution in linear algebra. Second, the Jacobian of $T_\KR$ is a lower triangular matrix and hence its determinant is given simply by 
\begin{equation} \label{eq:JacobianDeterminant}
|\nabla T_\KR(\w)| = \prod_{k=1}^d \partial_{w_k} T_k(\w_{1:k-1}, \w_k).
\end{equation}
In fact, this identity implies that $T_\KR$ is 
invertible whenever the 1D functions 
$T_k(\w_{1:k-1}, \cdot)$ are invertible for all values of $\w_{1:k-1}$. 
Finally, the formula for the Jacobian determinant of 
$T_\KR$ provides a tractable evaluation of the pushforward density $T_\KR \# \source_\W$ using the change-of-variables formula 
\begin{equation} \label{eq:ChangeOfVariables}
  T_\KR \# \source_\W(\z) = \source_\W(T_{\KR}^{-1}(\z))|\nabla T_{\KR}^{-1}(\z)|.  
\end{equation}

\rev{An important property of KR maps, which can be either 
beneficial or detrimental depending on applications, is that they inherently depend on an ordering of coordinates of the input variable $\w$; i.e., the maps are not invariant under changes of coordinates. For each ordering of the $\w_k$'s, the KR map is a uniquely defined monotone map, which follows from the uniqueness of each one-dimensional map in~\eqref{eq:OneDimMap}. When certain marginal conditionals of the target distribution $\target_\Z$ are of interest, the ordering can be chosen so that the corresponding components can be used to sample those conditionals. For estimation, however, different  orderings yield maps with different regularity. Moreover, some orderings lead to sparser or smoother maps, and hence can be approximated more efficiently; see~\cite{irons2022triangular, wang2022minimax, zech2022sparse, zech2022sparseb} for more details and the discussion in Section~\ref{sec:computation}.}
 

\subsection{Generalization to Block-triangular Maps} \label{sec:block-maps}

As motivated in~\Cref{sec:intro}, our goal is to condition a vector-valued random variable $\X \in \Xc$ on another variable $\Y \in \Yc$. Section~\ref{sec:KR_map} shows that a KR map with a careful choice of ordering for the target variables $(\Y, \X)$ 
can sample 
univariate marginal conditionals of the joint distribution for each coordinate. In this section, we introduce an extension of the KR map to directly sample the conditional distribution $\mu_{\X|\Y}$ for the multivariate variable $\X|\Y$. This construction is particularly advantageous when a natural ordering for the individual coordinates of $\X$ or $\Y$ may be unknown. 

Following \cite{baptista2024conditional}, we define a block-triangular map $T\colon \Yc \times \Xc \to \Yc \times \Xc$ as a 
mapping of the form  
\begin{equation} \label{eq:BlockTriangularMap}
T(v, u) =  \Big( T_\Yc(v), T_\Xc(v, u) \Big),
\end{equation}
with $\Yc$-component $T_\Yc\colon \Yc \to \Yc$ and 
 $\Xc$-component $T_\Xc\colon \Yc \times \Xc \to \Xc$ playing 
 the same role as the $T_k$ components for KR maps; 
 also recall that we used $(\v, \u)$ to denote 
 points in the domain of $T$ while $(\y, \x)$ denote points 
 in the range. 
We call these maps block-triangular because their Jacobians 
are block-triangular matrices. Note that we do not require any 
further structure on the $\Yc$ or $\Xc$ components and that 
the KR map, by virtue of being fully triangular, also belongs 
to the block-triangular family so long as we order 
$\w = (\v, \u)$, i.e., components of $\v$
appear before those of $\u$.

Let us consider a reference distribution $\source_{\V, \U}$ 
and a target $\target_{\Y, \X}$. Following the same 
ideas behind the construction of the KR map, we can 
take $T_\Yc$ as any map that transports the $\V$-reference marginal 
$\source_\V$ to the $\Y$-target marginal $\target_\Y$.
Such mappings exist under very general conditions as we will 
discuss in \Cref{sec:regularity}. Furthermore, 
we can  take $T_\Xc(\v, \cdot)$, viewed as a family of maps 
parameterized by $\v$, as the collection of maps transporting 
the reference conditionals $\source_{\U \mid \v}$ 
to the target conditionals $\target_{\X \mid T_\Yc(\v)}$. 
Once again such mappings exist under very general assumptions, for example 
when the reference conditionals $\source_{\U \mid v}$ have no atoms. 
We can verify directly that the block-triangular map that we just 
constructed satisfies $T \# \source_{\V, \U} = \target_{\Y, \X}$
and by construction 
$T_\Xc( \v, \cdot) \# \source_{\U \mid \v} 
= \target_{\X \mid T_\Yc(\v)}$.

It turns out that the conditioning property of 
block-triangular maps is not restricted to our construction above. 
In fact, any block-triangular map that transports 
$\source_{\V, \U}$ to $\target_{\Y, \X}$ can also map
their conditionals. This is the content of the following 
theorem borrowed from 
\cite{baptista2024conditional, hosseini2025conditional}.

\begin{theorem} \label{thm:triangular_conditional_pushforward} Let $T$ be any block-triangular  map satisfying 
$T \#  \source_{\V, \U} = \target_{\Y, \X}$. 
Then $T_\Xc( \v, \cdot) \# \source_{\U \mid \v} 
= \target_{\X \mid T_\Yc(\v)}$ for $v$ a.e.\ from $\source_\V$.
\end{theorem}
Just as in the case of KR maps, the above result is 
more useful in practice where we want to 
characterize $\target_{X \mid y}$.
Assuming $T_\Yc$ is invertible 
we can set $\v = T_\Yc^{-1}(\y)$ to obtain the 
identity $T_\Xc(T_\Yc^{-1}(\y), \cdot) \# \source_{\U \mid T_\Yc^{-1}(\y)}
= \target_{\X \mid \y}$ whenever $T \# \source = \target$.
Again, notice the similarity between this expression and \eqref{KR-map-component}.

Parallel to our first construction of the KR map, we can 
simplify our block-triangular maps by choosing a reference $\source_{\V, \U}$
that has a particular structure. Most notably, we can choose 
$\source_{\V, \U} = \target_\Y \otimes \source_\U$, i.e., the 
$\V$-marginal of the reference coincides with the target and 
$\V, \U$ are independent under the reference. In this case we can simply
take $T_\Yc = \text{Id}$ and work 
with block-triangular maps of the form 
\begin{equation}\label{block-triangular-map-simplified}
    T(y, u) = \Big( y, T_\Xc (y, u) \Big)
\end{equation}
which further yield the convenient identity 
\begin{equation}\label{eq:triangular-conditioning-simplified}
T_\Xc(\y, \cdot) \# \source_\U = \target_{\X \mid y},    
\end{equation}
for each $y \in \Yc$. As we will discuss later, these types of block-triangular
maps are particularly useful for Bayesian inference
and conditional generative modeling since, in many 
of those problems, we can often construct a reference $\source_{\V, \U}$
that has the desirable form  discussed above.

Finally, we note that the results outlined above 
have natural extensions to the case where 
$\Yc$ and $\Xc$ are abstract vector spaces such 
as function spaces or spaces of probability measures \cite{bogachev2005triangular,baptista2024conditional,hosseini2025conditional}. 
Such extensions have wide applications in scientific computing and uncertainty quantification, particularly in 
PDE inverse problems
\cite{stuart2010inverse}. One such extension where $\mathcal{\Y}$ is the space of probability distributions will be described  
in \Cref{sec:conditioning-dist-on-dist}. 




\subsection{Existence of Triangular Maps via Optimal Transport Theory}\label{sec:regularity}

So far in our discussions of block-triangular and KR maps, we 
primarily focused on their construction and key properties
related to the characterization of conditional distributions.
However, we did not establish the existence of 
block-triangular maps.
In this section we briefly discuss this topic 
and at the same time characterize block-triangular 
maps from the perspective of optimal transport \rev{(OT)}. 
 This connection is also very useful for developing 
 computational algorithms and error bounds.  
We will focus our 
discussion towards the block-triangular maps 
of \Cref{sec:block-maps} but note that the discussed 
ideas are deeply connected to KR maps \cite{carlier2010knothe, bonnotte2013knothe, baptista2024entropic}. 

\rev{OT} identifies transport maps $T$ that
minimize the average cost of displacing probability mass with respect to a cost function $c \colon \R^d \times \R^d \rightarrow \R$ via the Monge problem 
\begin{equation} \label{eq:MongeProblem}
  \min_{T} \left\{\left. \int c(\w,T(\w))d\source(\w) \right| T\# \source_\W = \target_\Z \right\}.
\end{equation}
The following celebrated Brenier theorem~\cite{brenier1991polar} 
(see also \cite[Theorem~1.22]{santambrogio2015optimal}), 
provides a solution to the Monge problem for the quadratic cost $c(\w,\w') = \frac{1}{2}\|\w-\w'\|^2$. 

\begin{theorem}\label{thm:Brenier-theorem}
Suppose $\source_\W, \target_\Z \in \P(\R^d)$ such that 
$\int |\w|^2 \dd \source_\W(\w), \int | \z|^2 \dd \target_\Z(\z) < +\infty$ and that 
$\source_\W$ assigns zero mass to sets of dimension less than or 
equal to $d-1$. Then there exists a unique 
\rev{OT} map $T \colon \R^d \to \R^d$ such that 
$T \# \source_\W = \target_\Z$ given by $T = \nabla \psi$ where 
$\psi \colon \R^d \to \R$ is a convex function.
\end{theorem}
We refer to the above \rev{OT} map as {\it the Brenier map}. 
It is important to note that while the Brenier map is unique, in general 
there may be infinitely many transport maps from 
$\source_\W$ to $\target_\Z$;
the Brenier map is unique among maps that are given as gradients of 
convex functions. We also note that Brenier maps are not triangular since they are gradients of convex functions, and hence have symmetric Jacobians. Regardless, we can use the above theorem to obtain the following existence result for block-triangular maps.

\begin{theorem}\label{thm:block-triangular-existence}
Consider reference and target distributions $\source_{\V, \U}$ and $\target_{\Y, \X}$ that satisfy the following conditions: 
\begin{itemize}
    \item The reference marginal $\source_{\V}$ assigns 
    zero mass to sets of dimension less than or equal to 
    $\dim(\Yc)$. 

    \item There exist reference conditionals 
    $\source_{\U \mid \v}$
    that assign zero mass to sets of dimension 
    less than or equal to 
    $\dim(\Xc)$ for all $\v \in \Yc$. 

    \item It holds that 
    \begin{equation*}
        \begin{aligned}
        & \int \|\v\|^2 + \|\u\|^2 \dd \source_{\V, \U}(v, u) < +\infty,\\
        & \int \|\y\|^2 + \|\x\|^2 \dd \target_{\Y, \X}(\y, \x) < +\infty.
        \end{aligned}
    \end{equation*}
\end{itemize}
Then there exists a unique block-triangular map 
$T_{\BT}(\v, \u) = \big(T_\Yc(\v), T_\Xc(\v, \u) \big)$ 
\rev{satisfying $T_{\BT}\# \source_{\V, \U} = \target_{\Y, \X}$}
whose components are of the form $T_\Yc(\v) = \nabla \psi(\v)$, 
for a convex function $\psi\colon \Yc \to \R$, and $T_\Xc(\v,\u) = 
\nabla_\u \phi(\v, \u)$, 
for a \rev{$\u$-convex} function $\phi\colon \Yc \times \Xc \to \R$.
\end{theorem}

\begin{proof}[Sketch of Proof]
We will outline the main ideas of the proof here and refer the reader to 
\cite{Carlier_2016,hosseini2025conditional, chemseddine2024conditional} for details. An application of \Cref{thm:Brenier-theorem}
with $\source_\W \leftarrow \source_\V$ and 
$\target_\Z \leftarrow \target_\Y$
yields the existence and uniqueness of $T_\Yc$. Next, we
apply Brenier's theorem to define the individual 
transport map $T_\Xc(\v, \cdot)$ as the 
Brenier map that pushes $\source_{\U \mid \v}$ 
to $\target_{X \mid T_\Yc(\v)}$ for each value of $\v$. 
These maps exist under the assumption that the reference conditionals $\source_{\U \mid \v}$ assign zero mass to 
low dimensional sets. 
Moreover, the map for each conditional is 
$T_\Xc(\v,\u) = \nabla_\u \phi(\v,\u)$
with $\phi$ being convex in $\u$. 
The main technical step is to 
``glue" together these Brenier maps for all conditionals. 
Due to the fact that 
conditional measures are unique only up to sets of measure zero, 
$\v \mapsto T_{\Xc}(\v, \cdot)$ is multi-valued and among all of its 
possible values one has to select one that leads to a measurable map overall. This kind of technical result 
is well-known in analysis under the term  {\it measurable selection} \cite[Ch.~18]{aliprantis2006infinite}. 
The articles cited above discuss this technicality in more detail.
\end{proof}

\rev{While the triangular map from~\Cref{thm:block-triangular-existence} is not an OT map in the standard sense, it is associated with a 
so-called conditional OT problem in the parlance of~\cite{hosseini2025conditional, chemseddine2024conditional}. Conditional OT maps are shown to be} the limit of 
a sequence of OT maps with a singular 
cost~\cite{hosseini2025conditional,baptista2024entropic}. 
\rev{
Consider maps 
\begin{equation}\label{eq:weighted-OT}
    T_\tau = \argmin_{T \; : \; T \# \source_\W = \target_\Z} 
    \int c_\tau(\w, T(\w)) \dd \source_\W(\w), 
\end{equation}
where we introduced the weighted cost 
\begin{equation*}
 c_\tau(\w, \w')  =  \| \w - \w' \|_{A(\tau)}^2
\end{equation*}
with }
$\| \w \|_A := 
\| A^{1/2} \w \|$ and the diagonal block-matrix $A(\tau) = \text{diag}(I_\Yc, \tau I_\Xc)$
    with $I_\Xc, I_\Yc$  
    denoting the identity maps on $\Xc, \Yc$. Then $T_\BT = \lim_{\tau \downarrow 0} T_\tau$ in $L^2(\source_\W)$ under suitable assumptions. In~\cite[Theorem~2.1]{baptista2024entropic}, it is  observed that $T_\tau$ takes the form $T_\tau(w) = A(\tau)^{-1/2} T_0(A(\tau)^{1/2}w)$ where $T_0$ is the Brenier map with respect to the standard quadratic cost from $A(\tau)^{1/2}\# \source_\W$ to $A(\tau)^{1/2}\# \target_\Z$. We note that $\nabla T_\tau$ is symmetric for all $\tau > 0$ while the limiting 
    map is not.

    \rev{
    The above construction can also be extended to fully triangular 
    transport maps  following~\cite{bonnotte2013knothe, carlier2010knothe}. 
    In particular, \cite[Theorem~2.2]{carlier2010knothe} shows that the 
    KR map can be obtained as the limit $\tau \downarrow 0$ of \eqref{eq:weighted-OT}
    with the choice of the matrix 
    $A(\tau) = \text{diag}( 1, \tau, \tau^2, \dots, \tau^{d} )$.
    }
    


\rev{Finally, we note that as $\tau \downarrow 0$, problem \eqref{eq:weighted-OT}
gives rise to a conditional OT problem and a corresponding conditional Wasserstein 
distance with analogous geometric properties and dynamic characterizations 
mirroring the standard Wasserstein distances~ \cite{chemseddine2024conditional}.
The connection between triangular maps and OT also leads to a number of algorithms and computational methodologies. Whether the OT map is the best target for practical computation, however, is a more subtle question that we will discuss in more detail  in~\Cref{sec:computation}.} 

\subsection{Regularity of Triangular Maps}

The purpose of this section is to bring together current results on the regularity of triangular maps. The results are central to the design of algorithms and the pertinent analyses of consistency, convergence, and estimation, but stand on their own and can be skipped by a familiar reader.

The \rev{OT} characterization of $T_\BT$ in
\Cref{thm:block-triangular-existence} reveals a path 
towards a first regularity analysis of block-triangular maps 
using the classic regularity result of Caffarelli~\cite{caffarelli1992regularity},
which we will recall in a simplified setting below.

\begin{theorem}\label{thm:caffarelli-regularity}
Let $\Omega \subset \R^d$ be  bounded and convex, and suppose 
that $\source_\W, \target_\Z \in \P(\Omega)$ have strictly 
positive $C^k(\Omega)$ densities 
with respect to the Lebesgue measure. Then the Brenier map $T$ pushing 
$\source_\W$ to $\target_\Z$ is $C^{k+1}(\Omega)$.
\end{theorem}

We can now combine this result with \Cref{thm:block-triangular-existence}
to obtain the following corollary that characterizes the regularity of the components of $T_\BT$. To state the result we need to introduce some notation: Given a  set $\Omega \subset \Yc \times \Xc$ we write 
$\Omega_\Xc:= \{ x \; : \; \exists \, y \text{ such that } 
(\y,\x) \in \Omega \}$ 
with $\Omega_\Yc$ defined in a similar manner. We also 
define the slices 
$\Omega_{\Xc \mid y} := \{ x \; : \; (\y,\x) \in \Omega \}$.

\begin{corollary}\label{cor:block-triangular-regularity}
Suppose that $\source_{\V, \U}, \target_{\Y, \X}$ 
are supported on a bounded and convex set $\Omega \subset \Yc \times \Xc$ 
and have $C^k(\Omega)$ densities that are strictly positive 
on this set. Furthermore, suppose that there exists a version of the 
conditionals $\source_{\U \mid \v}$ and $\target_{\X \mid \y}$
that are $C^k(\Omega_{\Xc \mid \y})$ for all $\v, \y \in \Omega_\Yc$. 
Then, the map $T_\BT$ exists with  components $T_\Yc \in C^{k+1}(\Omega_\Yc)$
and 
 $T_\Xc(\y, \cdot) \in C^{k+1}(\Omega_{\Xc \mid y})$ for all 
$\y \in \Omega_\Yc$.
\end{corollary}

\begin{proof}
    First, the boundedness of $\Omega$ and 
    the smoothness of $\source_{\V, \U}$ and $\target_{\X, \Y}$
    imply the conditions of \Cref{thm:block-triangular-existence} and give us the existence of $T_\BT$. Given that $\source_{\V, \U}, \target_{\X, \Y}$ are $C^k(\Omega)$, 
    then so are the marginals $\source_\V, \target_\Y$, and 
    an application of \Cref{thm:caffarelli-regularity}
    yields $T_\Yc \in C^{k+1}(\Omega_\Yc)$. The regularity of 
    the components $T_\Xc(v,\cdot)$ follows in a similar 
    manner by applying \Cref{thm:caffarelli-regularity}
    to the Brenier maps transporting $\source_{\U \mid v}$ 
    to $\target_{\X \mid T_\Yc(v)}$.
\end{proof}

It is important to note that the above regularity result is 
in some sense incomplete. More precisely, the corollary does not 
provide $C^k$ regularity of the overall map $T_\BT$. This is indeed 
a challenging theoretical question which, to our knowledge, is 
open for the general case of block-triangular maps. The main 
difficulty is in understanding the regularity of $T_\Xc$ with respect
to the $\Yc$-coordinate. To do so, one can use that the conditionals $\source_{\U \mid \v}$ and  $\target_{\X \mid T_\Yc(\v)}$ vary smoothly with $\v$ and combine it with quantitative estimates on the variation of the $T_\Xc(\v, \cdot)$ maps. This amounts to obtaining quantitative stability estimates for the solutions of the Monge--Amp{\`e}re PDEs associated with the transport from $\source_{\U \mid \v}$ to $\target_{\X \mid T_\Yc(\v)}$; see \cite{figalli2017monge} for a discussion on the Monge--Amp{\`e}re PDE and its connection to Brenier maps and their regularity.
 \rev{
 In the simplified setting where 
 $\source_{\V,\U} = \target_\Y \otimes \source_\U$, 
 this line of thinking gives a H\"older regularity 
 result for the $T_\Xc$ map. The following theorem is a 
 special case of \cite[Corollary~4.1]{gonzalez2024linearization}.
 }

\begin{theorem}\label{cor:block-triangular-Holder-continuity}
\rev{    
    Suppose $\Omega_\Xc \subset \R^d$ is  bounded and uniformly convex, and suppose that 
    the conditionals $\target_{X \mid y}$ have strictly positive densities
    on $\Omega_\Xc$ for every 
    $y$ in the interior of $\Omega_\Yc$. Furthermore, suppose that   
    $\log \mu_{X \mid Y}(x \mid y) \in C^{1, \gamma}(\Omega)$ for some H\"older exponent $\gamma \in (0, 1]$ and  
     $\log \source_\U$ is $C^\alpha(\overline{\Omega_\Xc})$ for some  H\"older exponent $\alpha \in (0,\gamma)$.
    Then $y \mapsto T_\Xc(y, \cdot) \in C^{1, \alpha}(\overline{\Omega_\Xc})$ is 
    $C^1$.
}
\end{theorem}

\rev{
Finally, a more direct and classical argument can characterize the regularity of 
strictly-triangular KR maps. In particular, the explicit form of the KR map components~\eqref{KR-map-component} in terms of the conditional CDFs of the reference and target distributions implies that the KR map inherits an extra degree of regularity from the reference and target's marginal conditional densities. As stated in~\cite[Remark~2.19]{santambrogio2015optimal}, it 
also follows that the KR map components $T_k$ are regular with 
respect to $w_1, \dots, w_k$ although this regularity can be 
anisotropic in 
general: For example, $T_k$ typically has $s+1$ partial derivatives 
with respect to $w_k$ when it has $s$ partial derivatives with 
respect to $w_1, \dots, w_{k-1}$. Additionally, the ordering of 
the variables has a major impact on the regularity of the $T_k$
components which in turn affects approximation rates of 
algorithms; see \cite{irons2022triangular, wang2022minimax}
as well as our discussion of estimation rates in \Cref{sec:estimation-rates}.
}

\section{Estimation of triangular maps} \label{sec:computation}

\rev{
In this section we will introduce and discuss practical estimators for triangular transport maps. In~\Cref{sec:objectives} we identify triangular transports by solving a variational problem over a family of maps that is amenable for computation. In~\Cref{sec:error-analysis} we present known statistical results for some of these estimators.}

\subsection{Variational Characterizations of Maps} \label{sec:objectives}

\rev{In this section we present approaches to seek the triangular maps $T$ as the solution of an optimization problem:
$$\min_{T \in \mathcal{T}} \mathsf{L}(T,\mu_Z,\source_W),$$
or its empirical approximation over a space of transport maps $\mathcal{T}$, e.g., triangular or block-triangular maps of the form in~\eqref{eq:KR} or~\eqref{eq:BlockTriangularMap}. We write the loss $\mathsf{L}$ here in its most general form and note that the following subsections will introduce different loss functions for identifying $T$ that 
encompass most existing practical algorithms. These include statistical divergences in~\cref{sec:divergence_estimators},  OT objectives in~\cref{sec:optimal_transport}, and losses for dynamic transport in~\cref{sec:dynamic_transport}. We recall our notation for the concatenated variables $\Z = (\Y,\X)$ and emphasize that in this section 
all losses are evaluated on  $\Zc = (\Yc \times \Xc)$.}

\subsubsection{Minimum Divergence Estimators} \label{sec:divergence_estimators}
To approximately enforce the pushforward constraint $T \# \source_{\W} = \target_{\Z}$, minimum divergence estimators minimize a distance between the distributions $\target_{\Z}$ and $T \# \source_{\W}$. Let $\D \colon \mathcal{P}(\mathcal{\Zc}) \times \mathcal{P}(\mathcal{\Zc}) \rightarrow \R$ be a divergence between probability distributions. 
Then, consider the optimization problem
\begin{equation*}
\min_{T \in \mathcal{T}} \D(\target_{\Z}, T \# \source_{\W}).
\end{equation*}
While the choice of $\D$ is 
mostly problem dependent, we will outline three commonly used divergences.  

The Kullback--Leibler (KL) divergence from $\nu_2$ to $\nu_1$ is 
$$\dkl(\nu_1||\nu_2) = \int \log \frac{\dd \nu_1}{\dd \nu_2} \dd\nu_1.$$ Minimizing the KL divergence from the pushforward 
$T \# \source_{\W}$ to $\target_{\Z}$ amounts to solving 
\begin{equation} \label{eq:MLE}
\min_{T \in \mathcal{T}} \int -\log T \#  \source_\W(\z) \dd \target_\Z(\z),
\end{equation}
which can be made empirical given  samples from $\target_\Z$. This corresponds to solving a maximum likelihood problem 
for the target samples with respect to the class of pushforward distributions defined by the family $\mathcal{T}$. 

If the class $\mathcal{T}$ consists of invertible transformations (see Section~\ref{sec:parameterization} for more details), we can use the change-of-variables formula in~\eqref{eq:ChangeOfVariables} to express the pushforward density $T \# \source_\W$ in~\eqref{eq:MLE}. 
For log-concave reference distributions and (monotone) triangular maps, \eqref{eq:MLE} is then a convex optimization problem with respect to the inverse map $T^{-1}$ for any target distribution. 
Furthermore, when $\mathcal{T}$ is a family of strictly triangular maps as in~\eqref{eq:KR} and $\source_\W$ is i.i.d., \cite{marzouk2016sampling} shows that the optimization problem~\eqref{eq:MLE} decouples over the map components $T_k$. \rev{The following remark reveals that learning block-triangular maps using~\eqref{eq:MLE} is a generalization of many supervised learning problems.}

\begin{remark} \label{rem:supervised_learning} 
Let the reference distribution be $\source_{\V,\U} = \target_\Y \otimes \source_{\U}$ where $\source_\U$ is a standard Gaussian distribution and $T$ be a block-triangular map of the form
\begin{equation} \label{eq:AffineMaps}
    T(\y,\u) = (Id(\y), m(\y) + \u),
\end{equation} 
where $m \colon \mathcal{\Y} \rightarrow \mathcal{\X}$. By construction, $T$ is invertible and the pushforward distribution is given by $T\#\source_\W(z) \propto \exp(-|T^{-1}(z)|^2)$. Hence, \eqref{eq:MLE} reduces to solving 
$$\min_{m} \int |\x - m(\y)|^2 \target_{\Y,\X}(\y,\x)\dd\y\dd\x.$$
This corresponds to finding the conditional expectation of the conditional distribution $\target_{\X|\Y}$, i.e., the minimizer for $m$ over all measurable functions is $m(y) = \mathbb{E}[\X|\Y = \y]$. \rev{Thus, 
learning a block-triangular map allows us to 
learn the entire conditional distribution rather than the 
conditional mean. }
\end{remark}

A second divergence $\D$ is the maximum mean discrepancy (MMD)~\cite{gretton2012kernel}.  
Let $k \colon \mathcal{\Z} \times \mathcal{\Z} \rightarrow \R$ be a symmetric positive definite  kernel function which induces a Reproducing Kernel Hilbert space (RKHS). Then, the squared MMD between 
$\nu_1$ and $\nu_2$ is 
\begin{align*}
\D_{\MMD}^2(\nu_1,\nu_2) &= \mathbb{E}_{(z_1,z_1') \sim \nu_1 \otimes \nu_1} k(z_1,z_1') \\
&\quad - 2\mathbb{E}_{(z_1,z_2) \sim \nu_1 \otimes \nu_2} k(z_1,z_2) \nonumber \\
&\quad + \mathbb{E}_{(z_2,z_2') \sim \nu_2 \otimes \nu_2} k(z_2,z_2'). \nonumber
\end{align*}
Given independent batches of samples from the two distributions $\nu_1,\nu_2$, the MMD can be estimated empirically \cite{sriperumbudur2012empirical}, sometimes defined as the empirical MMD distance. This objective was used in~\cite{pandey2024diffeomorphic} to seek transport maps, which were also taken to be in a RKHS. We also refer to~\cite{de2023diffeomorphic} for a related objective to MMD known as the Sinkhorn divergence, which selects the kernel $k$ based on an entropic \rev{OT} cost. 

The MMD is a special case of so-called integral probability metrics (IPMs) that compare expectations between two distributions~\cite{muller1997integral}. Given a class $\mathcal{F}$ of functions $f \colon \Zc \rightarrow \R$, known as a \emph{discriminator}, an IPM between two distributions is given by
$$\mathsf{D}_{\textrm{IPM}}(\nu_1,\nu_2) = \sup_{f \in \mathcal{F}} \bigg| \int f(\z)\dd\nu_1(\z) - \int f(\z)\dd\nu_2(\z) \bigg|.$$
Different choices for the discriminator class $\mathcal{F}$ result in different metrics. For example, choosing $\mathcal{F}$ to be Lipschitz-1 functions yields the dual-form of the Wasserstein-1 distance. The latter distance was popularized in~\cite{goodfellow2014generative, arjovsky2017wasserstein} to learn transport maps, known as generative adversarial networks (GANs), by jointly optimizing $T$, often called a generator, and the discriminator $f$. As compared to the KL divergence, IPMs do not require evaluating the pushforward density, and in particular the map's Jacobian $\nabla T$. As a result, these metrics are applicable even when the map is non-invertible. Certain generative models, such as GANs and variational auto-encoders, seek a map $T \colon \R^{r} \rightarrow \R^d$ for some $r \ll d$ that can identify a low-dimensional embedding of the target distribution. This approach was extended to conditional simulation using block-triangular maps in~\cite{baptista2024conditional}.

\subsubsection{Optimal Transport Estimators} \label{sec:optimal_transport}

Consider the entropic regularized Kantorovich relaxation of the OT problem in~\eqref{eq:weighted-OT}, which seeks a coupling $\gamma \in \mathcal{P}(\Zc \times \Zc)$, known as a \emph{transport plan}, by solving
\begin{equation} \label{eq:EntropicOTProblem}
    \min_{\gamma \in \Pi(\source_\W,\target_{\Z})} \int c_\tau(z,z') \dd\gamma(z,z') + \varepsilon \dkl(\gamma||\source_\W \otimes \target_\Z),
\end{equation}
where $\varepsilon > 0$ is a regularization parameter and $\Pi(\source_\W,\target_\Z)$ is the set of all plans that have marginals $\source_{\W}$ and $\target_{\Z}$. As $\varepsilon\rightarrow 0$, this reduces to~\eqref{eq:weighted-OT}.
For $\varepsilon, \tau > 0$, \eqref{eq:EntropicOTProblem} is strictly convex with a unique minimizer, \rev{which in terms of dual potentials $f_{\tau,\varepsilon} \in L^1(\source_\W)$, $g_{\tau,\varepsilon} \in L^1(\target_\Z)$ for the marginal constraints is} 
$$\frac{\dd\gamma_{\tau, \varepsilon}}{\dd(\source_\W\otimes\target_\Z)}(z,z') = \exp(f_{\tau,\varepsilon}(z) + g_{\tau, \varepsilon}(z') - c_\tau(z,z')/\varepsilon).$$
When $\source_\W,\target_\Z$ are empirical distributions each with $N$ samples, the dual variables are vectors in $\R^N$ that can be computed efficiently via Sinkhorn iterations~\cite{peyre2019computational}.

\rev{One may then estimate $T_\BT$ for fixed $\tau,\varepsilon$ by the \emph{entropic map}, defined as the conditional expectation} $\mathbb{E}_{\gamma_{\tau,\varepsilon}}[Z'|Z = \z]$, which 
has the explicit form
$$T_{\tau,\varepsilon}(\z) \coloneqq \int \frac{ \z' \exp(g_{\tau,\varepsilon}(\z') - c_\tau(\z,\z')/\varepsilon)}{\int \exp(g_{\tau,\varepsilon}(\z'') - c_\tau(\z,\z'')/\varepsilon) d\mu_\Z(\z'')}d\mu_\Z(\z').$$
This map can be made empirical given samples from $\target_\Z$ and a Sinkhorn estimate of $g_{\tau,\varepsilon}$. While the discrete OT problem only defines the plan at the samples, $T_{\tau,\varepsilon}$ can be evaluated at any $\z \in \Zc$, and \cite[Theorem~4.2]{baptista2024entropic} provides conditions under which it converges to \rev{$T_\BT$} in $L^2(\source_\W)$ as $\tau,\varepsilon \downarrow 0$.
\rev{We note that as compared to maps with a strict block-triangular structure, however, the entropic map for $\tau,\varepsilon > 0$ is only approximately block-triangular. In practice, the resulting maps can still yield adequate conditional samples.}

\subsubsection{Dynamic Transport and Flow Matching} \label{sec:dynamic_transport}
\rev{Rather than directly defining transport maps, dynamic approaches define $T$ implicitly via the flow map of an ordinary differential equation (ODE)~\cite{chen2018neural}. 
Given a velocity field $h \colon \mathcal{\Z} \times [0,1] \rightarrow \mathcal{\Z}$, these approaches evolve the ODE  
\begin{equation} \label{eq:NeuralODE}
\frac{dz}{dt} = h(z(t),t), \qquad z(0) = z_0,
\end{equation}
to end time $t = 1$ so that $T(z_0) = z(1) \sim \target_{\Z}$ for $z_0 \sim \source_{\W}$. By choosing $h$ to have the block-triangular form in~\eqref{eq:BlockTriangularMap}, $T$ is also guaranteed to be a block-triangular map~\cite{wang2023efficient}.}

\rev{One approach to find the dynamic transport~\eqref{eq:NeuralODE} is to minimize a statistical divergence as in~\Cref{sec:divergence_estimators}. However, this problem is ill-posed because there are many paths of time-dependent distributions that evolve from the reference $\source_\W$ to the target $\target_\Z$. To address this, some approaches seek the velocity field $h$ that also minimizes an OT penalty~\cite{onken2021ot}
or a regularization term on $h$ \cite{pandey2024diffeomorphic}. In the same way that OT selects one transport in a possibly infinite set of maps that satisfy the pushforward constraint, an alternative recent approach, known as \emph{flow matching}, selects one path whose velocity field can be 
learned using supervised learning techniques; see~\cite{lipmanflow, albergo2024stochastic} for more details and~\cite{liu2023flow, hertrich2026relation} for  connections to OT.}

\rev{One popular path is defined by the linear interpolant\footnote{A more general interpolant that is nonlinear with respect to time $t$ and also depends on an independent random variable is proposed in~\cite{albergobuilding}} 
\begin{equation} \label{eq:Interpolant}
\Z_t = (1 - t) \Z_0 + t \Z_1,
\end{equation}
between the random variables $Z_0 \sim \source_\W$ and $Z_1 \sim \target_\Z.$ \cite[Theorem~3]{lipmanflow} shows that taking the velocity field in~\eqref{eq:NeuralODE} to be the conditional expectation $h(z,t) = \mathbb{E}[Z_1 - Z_0|Z_t = z]$ yields a transport where the marginal density of $z(t)$ solving~\eqref{eq:NeuralODE} 
follows the same law as~\eqref{eq:Interpolant} for each time $t$. It can be easily shown that this velocity is 
the minimizer of the following problem
$$\min_{h \in \mathcal{H}} \int_0^1 \mathbb{E}_{(\z_0,\z_1)} \|\z_1 - \z_0 - h((1-t)\z_0 + t\z_1,t)\|^2 \dd t,$$
where $\mathcal{H}$ is a family of parameterized block-triangular velocity fields and the expectation is taken over a joint distribution of $(\Z_0,\Z_1),$ such as a product coupling, $\source_\W \otimes \target_\Z$, or an (approximate) \rev{OT} coupling~\cite{tong2024improving}. In~\cite{kerrigan2024dynamic}, the authors used \rev{OT} problem~\eqref{eq:EntropicOTProblem} to define the joint distribution. We also refer the reader to~\cite{batzolis2021conditional, dasgupta2025conditional} for related approaches based on score-based conditional diffusion models, a cousin of flow matching methods where the reference distribution is chosen to be a multivariate Gaussian.} 

\subsubsection{\rev{Implementation Details}} \label{sec:parameterization}

\rev{Implementing the estimators described above requires several choices beyond the empirical approximations of objectives. In this section we discuss the parameterization of the transport map as well as the numerical evaluation and inversion of these maps where pertinent. We argue that these choices are often coupled: for example, likelihood-based objectives require an invertible map with a tractable Jacobian determinant, while IPMs can accommodate non-invertible maps.
}

\paragraph*{\rev{Map Parameterizations}} \rev{
The first choice is the parametric class $\mathcal{T}$ over which the transport is learned. This architecture should directly enforce the dependence structure for the block-triangular map~\eqref{eq:BlockTriangularMap}, where the first component $T_{\Yc}$ depends only on $\v$, while the second component $T_{\Xc}$ depends on both $\v$ and $\u$. Similarly, for the strictly-triangular map~\eqref{eq:KR}, the $k$th component is restricted to depend only on the first $k$ input variables. In practice, the triangular architecture will require an ordering of variables; see Section~\ref{sec:error-analysis} for a discussion on optimal ordering. We now discuss three parametric classes for the map's components.}

\rev{A popular class for $\mathcal{T}$ is that of normalizing flows, which define each $T \in \mathcal{T}$ as the composition of simple invertible transformations, often parameterized by neural networks. The transformations are selected so that their inverse and Jacobian determinant~\eqref{eq:JacobianDeterminant} can be computed efficiently~\cite{rezende2015variational, tabak2013family, papamakarios2021flows, kobyzev2020normalizing}. These properties are particularly useful for minimization with the KL objective in~\eqref{eq:MLE}. In particular, autoregressive and coupling flows~\cite{dinh2016density, papamakarios2017masked, huang2018neural, de2020block} compose maps of the form in~\eqref{eq:AffineMaps} with variable permutations between consecutive maps. The permutations increase the map's overall expressiveness and reduce its sensitivity to variable ordering. For conditional simulation, however, it is important that compositions preserve block-triangular structure (as in Theorem~\ref{thm:block-triangular-existence}). This can be achieved by permuting variables only within the prescribed blocks.}

\rev{A class of strictly triangular maps parameterizes  each component $T_k$ in~\eqref{eq:KR} so that monotonicity in its final argument is guaranteed, 
mimicking the theoretical properties of KR maps. 
For example, one may write
\[
T_k(\w_{1:k})
=
a_k(\w_{1:k-1})
+
\int_0^{w_k}
g\!\left(
b_k(\w_{1:k-1},t)
\right)\,dt,
\]
where $a_k$ and $b_k$ are unconstrained functions and $g\colon\R\rightarrow\R_{+}$ is a positive link function, such as $g(t)=\exp(t)$ or $g(t)=\log(1+e^t)$. Then, $\partial_{w_k}T_k(\w_{1:k}) =
g\!\left(b_k(\w_{1:k})\right)>0,$
and hence $w_k\mapsto T_k(\w_{1:k})$ is strictly increasing for fixed $\w_{1:k-1}$. This form converts a constrained optimization problem over monotone functions $T_k$ into an unconstrained problem over $a_k$ and $b_k$, which are often represented using polynomials~\cite{baptista2024representation} or neural networks~\cite{wehenkel2019unconstrained}. For certain choices of the link function and approximation space,~\cite{baptista2024representation} establishes favorable properties of the resulting optimization problem.
We note that this integral representation introduces some numerical considerations, i.e., evaluating the map may require numerical quadrature.
We also refer to a related approach~\cite{cui2022deep, cui2023scalable} that maps from the target distribution to a uniform reference $\source = \mathcal{U}[0,1]^d$ by parameterizing the CDFs of the marginal conditional target distributions using tensor-train approximations.}

\rev{A third parameterization follows from Brenier's characterization of quadratic-cost \rev{OT} maps as gradients of a convex potential function $\psi \colon \Zc \rightarrow \R$, i.e.,
\[
T(\w)=\nabla\psi(\w).
\]
The potential $\psi$ can be approximated using input-convex neural networks (ICNNs)~\cite{amos2017input}, which enforce convexity through architectural restrictions and constraints on the network weights. For block-structured conditional transports, one may instead impose convexity only in the variables being transported. In particular, parameterizations of the form $T_{\Xc}(\v,\u)=\nabla_{\u}\psi(\v,\u)$, where $\psi(\v,\cdot)$ is convex for every fixed $\v$, can be implemented using partial-ICNN architectures~\cite{bunne2022supervised, wang2023efficient}. These parameterizations directly enforce monotonicity and hence invertibility of each block, although evaluating the inverse often requires solving a convex optimization problem.}

\rev{Finally, an alternative to enforcing the invertibility in each element of $\mathcal{T}$ is to enforce invertibility approximately during optimization. This can be achieved by adding a regularization term to the objective. For example, \cite{baptista2024conditional} uses a regularizer that promotes monotonicity of the map in expectation over all inputs, while \cite{hosseini2025conditional} uses an \rev{OT} penalty that biases the optimization toward the optimal and monotone Brenier map. 
Although the resulting maps are not guaranteed to be invertible everywhere, 
they can be used with flexible architectures which are easier to implement in practice. For example, we can use these with GANs that parameterize the map as $T \colon \R^{r} \rightarrow \R^d$ for some $r \ll d$. This is particularly useful when training with objectives that do not require evaluating the target density, such as IPMs.}


\paragraph*{\rev{Map Evaluation}}

\rev{
The computational cost of using a triangular transport map depends on the choice of the parameterization. For a monotone strictly triangular map of the form in~\eqref{eq:BlockTriangularMap}, sampling the target distribution is performed by evaluating each component in parallel. On the other hand, inverting the map is performed sequentially by solving 
a series of root finding problems and forward-substitution: 
solve for $w_1$, then solve for $w_2$ given $w_1$, and so forth.
When the map is implicitly defined by an ODE~\eqref{eq:NeuralODE}, its numerical evaluation or inversion also requires selecting an ODE solver, step size, error tolerances, and other parameters, 
taking into account the usual issues that arise in integration of 
high-dimensional ODEs.
Numerical integration also produces additional errors that may be 
absent for other types fo maps.
}

\rev{Finally, we note that a degree of freedom in practice is the direction of the approximated map, i.e., whether we parameterize the map $T$ or its inverse $T^{-1}$ directly. This choice often depends on whether the dominant downstream use is density evaluation or conditional sampling. A map $T$ pushing forward the reference to the target distribution is convenient for generating target samples, whereas the inverse map $T^{-1}$ is convenient for evaluating the target density using~\eqref{eq:ChangeOfVariables}. The latter is particularly useful when solving \eqref{eq:MLE}.
}


\subsection{\rev{Statistical Error Analysis}} \label{sec:error-analysis}

\rev{While there is now substantial statistical theory for estimating OT maps from discrete or continuous data (see~\cite{balakrishnan2025statistical} for a recent overview), corresponding results for triangular transports remain comparatively limited. In this section we outline existing results that address two related questions: estimation of the transport map and approximation of the target distribution.}

\subsubsection{Transport Map Estimation} \label{sec:estimation-rates}
\rev{Let $S_{\KR}=T_{\KR}^{-1}$ denote the monotone triangular map from the target measure $\mu_{\Z}$ to the reference measure $\source_\W$, and let $\widehat S_N$ be a maximizer of an empirical estimate for the log-likelihood
in~\eqref{eq:MLE} given $N$ i.i.d.\ samples over a $s$-smooth class of triangular maps. Under suitable assumptions on the reference and target densities, i.e., the densities being bounded away from zero, having compact convex support, and uniformly bounded derivatives up to order $s$, and 
the reference being strongly log-concave, 
\cite[Theorem~4.1]{irons2022triangular} gives
\begin{equation*}
 \mathbb{E}\!\left[
   \|\widehat S_N-S_{\KR}\|_{H^{1,2}(\target_{\Z})}^{2}
 \right]
 \lesssim
 \begin{cases}
   N^{-1/2}, & d<2s,\\
   N^{-1/2}\log N, & d=2s,\\
   N^{-s/d}, & d>2s.
 \end{cases}
\end{equation*}
Here, the Sobolev norm $H^{1,2}(\target_\Z)$ controls both the map and its derivatives. The same rates hold for the inverse estimator
$\widehat T_N=\widehat S_N^{-1}$ in $H^{1,2}(\source_\W)$; see the supplementary
results of \cite{irons2022triangular}. These bounds follow from
empirical-process-theoretic control of the log-likelihood together with strong convexity of
the objective for a strongly log-concave reference distribution.} 

\rev{The analysis in \cite{irons2022triangular} also accounts for anisotropic
smoothness. If the target density has coordinate-wise smoothness of orders
$(s_1,\ldots,s_d)$, the rate above depends on the effective smoothness of the marginal conditional densities that define the KR map. Consequently, 
the analysis 
suggests that ordering the variables from least to most smooth, i.e., 
$s_1\leq\cdots\leq s_d$ minimizes the estimation error.  However, we note that this oracle result does not account for the statistical cost of identifying an ordering from data which 
is crucial in practice.}

\rev{Sharper estimation rates are also available for the resulting pushforward distribution of a nonparametric estimator for the map. 
For target distributions $\mu_\Z$ with a $C^\alpha$-smooth density that is 
uniformly bounded above and away from zero, 
Proposition~2.1 of
\cite{wang2022minimax} shows that the corresponding KR map belongs to an
anisotropic H\"older class. 
Then for $\alpha>d/2$, Theorem~2.3 of
\cite{wang2022minimax} shows that the
maximum-likelihood estimator $\widehat{T}_N$ satisfies the following rates in the Hellinger distance $\mathsf{D}_H$ and the KL divergence: 
\begin{align*}
 &\mathsf{D}_H\!\left(\mu_\Z, (\widehat T_N)\#\source_\W \right)
   = \mathcal{O}_{\mathbb{P}}\!\left(N^{-\alpha/(2\alpha+d)}\right) \\
 &\dkl\!\left(
   \mu_\Z \,\middle\|\,(\widehat T_N)\#\source_\W
 \right)
   = \mathcal{O}_{\mathbb{P}}\!\left(N^{-2\alpha/(2\alpha+d)}\right).
\end{align*}
These rates match the classical minimax-optimal rates for estimating an
$\alpha$-smooth density in dimension $d$. Analogous minimax-optimal results
are also established for penalized likelihood estimators with Sobolev-type
penalties. 
Minimax optimality for the induced density does not, however, imply minimax optimality for the transport map, i.e., matching upper and lower bounds for an estimated KR map are not currently available.}

\rev{Similar statistical learning theory has also been developed for dynamic transport maps; recall Section~\ref{sec:dynamic_transport}.
The article \cite{JMLR:v25:23-1280} analyzes nonparametric maximum-likelihood
estimators over classes of time-dependent ODE velocity fields~\eqref{eq:NeuralODE}. The construction uses the KR map to exhibit a sufficiently regular admissible velocity field. 
For target distributions with $C^k$ smooth densities on the unit cube and 
$C^k$ velocity classes, the authors obtain for some $0<\gamma<k-\frac {d}{2}-\frac{3}{2}$, the following rate in Hellinger distance:
\begin{equation*}
 \mathbb{E}\!\left[
 \mathsf{D}_H\!\left(\target_\Z, (\widehat{T}_N)\#\source_\W, \right)
 \right]
 \lesssim
 N^{-\frac{2(k-1-\gamma)}
              {2(k-1-\gamma)+d+1}}.
\end{equation*}
The authors also obtain nearly minimax-optimal rates for neural-network velocity
fields. The appearance of $d+1$, rather than $d$, reflects the additional
time variable in the velocity field. We also refer the reader to~\cite{fukumizu2025flow} for related analyses of flow matching models without triangular structure which obtain nearly minimax-optimal
distribution-estimation rates in $p$-Wasserstein distance.}

\subsubsection{Conditional Density Estimation} \rev{While the preceding results concern estimation of a complete KR map or the joint pushforward density, they do not  quantify the accuracy of the conditional distributions obtained from some components of a block-triangular map. In particular, convergence in the joint distribution does not directly provide convergence guarantees for conditional distributions either uniformly or pointwise over the conditioning variable. General results on conditional density estimation have  derived minimax estimation rates under total-variation smoothness~\cite{li2022minimax} 
and through empirical entropy~\cite{bilodeau2023minimax}. These results are not specific to transport estimators, but describe rates that the pushforward distribution of triangular maps might seek to attain.}

\rev{One statistical result for conditional distributions is provided by recent analyses of conditional diffusion models~\cite{batzolis2021conditional}. Let $\X\in\mathbb{R}^{d_\X}$ denote a response
and $\Y\in\mathbb{R}^{d_\Y}$ a conditioning variable, and suppose that the
conditional distribution $\mu_{\X \mid \Y}$ has density with smoothness $\alpha_\Y$ in $\y$ and
$\alpha_\X$ in $\x$. Up to logarithmic factors, 
\cite{tang2025conditional} establishes that an approximate conditional distribution $\widehat{\target}_{\X \mid \Y}$ found with $N$ target samples satisfies
\begin{equation*}
 \mathbb{E}_{\Y}\!\left[
   d_{\mathrm{TV}}\!\left(
      \mu_{\X \mid \Y}, \widehat\mu_{\X \mid \Y}
   \right)
 \right]
 \lesssim
 N^{-\left(2+d_X/\alpha_X+d_Y/\alpha_Y\right)^{-1}}.
\end{equation*}
In addition, the authors show that this rate is minimax-optimal in the regimes treated therein and also obtain a similar rate in Wasserstein-1 distance. Extensions of the result can replace the ambient dimensions by intrinsic dimensions when the
conditioning variables and conditional distributions possess
low-dimensional structure. An interesting direction for future work is to obtain comparable minimax rates and guarantees adapting to intrinsic dimensions for block-triangular transport maps.}

\rev{The assumed tail behavior of the reference and target distributions also determines whether a
regular triangular model is correctly specified. \cite{jaini2020tails} show
that a globally Lipschitz transport cannot map a light-tailed reference
distribution to a target with strictly heavier tails. More precisely, the
source and target quantile-density functions determine the asymptotic growth of the one-dimensional monotone transports~\eqref{eq:OneDimMap}. In other words, the map component and its derivatives may be unbounded for reference and target distributions with mismatched tails. Consequently, estimators based on Lipschitz or bounded-derivative maps can remain misspecified even as the sample size and approximation-space
dimensions grow. Most of the available converges rates described above, which are primarily proved on compact domains, do not cover this setting.}

\rev{Finally, we distinguish approximation theory results from the preceding statistical guarantees. For analytic, strictly positive densities on $[-1,1]^d$, \cite{zech2022sparse} prove that the KR map and its inverse admit monotone triangular approximations based on sparse polynomials or deep ReLU networks. For an approximation space of dimension $M$, the induced distributional errors in Hellinger, total variation, Wasserstein, and KL, are shown to decay as $\exp(-\beta M^{1/d})$ for sparse-polynomial approximations and as $\exp(-\beta M^{1/(d+1)})$ for a neural-network construction. These rates are exponential in the approximation level but retain a dimension-dependent exponent. 
Under analyticity assumptions, \cite{zech2022sparseb} 
extends these regularity and approximation results to the infinite-dimensional setting with direct applications to 
Bayesian inverse problems for PDEs with function-space unknowns; see also~\Cref{sec:extended_methodology} for more discussion on this. We note that these approximation results do not account for the additional statistical error incurred when a map is learned from samples. 
Finally, to address the dependence of the KR map's regularity  on the chosen coordinate ordering, \cite{zech2022sparse} propose a criterion for selecting the ordering that minimizes approximation error.}



\section{Conditional Generative Modeling and Bayesian Inference} \label{sec:conditional_modeling}

In this section we describe how triangular transport maps are used for conditional generative modeling and Bayesian inference. \Cref{sec:SBI} considers the setting in which samples from the joint distribution $\target_{\Y,\X}$ are available to learn a block-triangular map, which can characterize the conditional distribution $\target_{\X|\y}$ at a new observation. Then, \Cref{sec:composed_maps} shows how composing a learned map with its partial inverse can improve the quality of conditional sampling when the estimated map is imperfect. We also relate this construction to residual resampling and regression adjustment. 

\subsection{Simulation-Based Inference} \label{sec:SBI}

As described in~\Cref{sec:BIP}, some Bayesian inverse problems feature latent variables and result in likelihood functions or prior densities that are intractable to evaluate. Without access to the likelihood or prior we cannot employ Bayes' rule in the usual sense. Instead, simulation-based methods perform inference in these settings by using paired data samples of parameters and observations. These samples are typically obtained by sampling a parameter from the prior distribution $\target_\X$ and an observation from the conditional distribution $\target_{\Y|\X}$. That is, 
$$\x^{(i)} \sim \target_\X, \qquad \y^{(i)} \sim \target_{\Y|\x^{(i)}}.$$ 
The resulting pairs $(\y^{(i)},\x^{(i)})$ are samples from the joint distribution $\target_{\Y,\X}$. Our goal is to sample from the conditional distribution $\target_{\X|\y}$ for a specific observation $\y \in \Yc$ given multiple i.i.d.\ sample pairs.

Observe that each pair $(\y^{(i)},x^{(i)})$ consists of a conditional sample $\x^{(i)} \sim \target_{\X|\Y}(\cdot|\y^{(i)})$ for the observation $\y^{(i)}$. In practice, we will only see a single sample from each conditional distribution and none of the $\y^{(i)}$ samples will match the conditional value $\y \in \Yc$. Thus, the goal is only tractable by learning a map that can sample from any conditional distribution $\target_{\X|\y}$ given $y \in \Yc$. We follow the methodology in Section~\ref{sec:objectives} to learn a block-triangular map $T_{\BT}$~\eqref{eq:BlockTriangularMap} which achieves this goal. Learning in this context is also known as \emph{amortized inference} because a single map is used to solve multiple inverse problems for different observations rather than solving a new inference problem for every observation (as in common in MCMC). Here, one typically seeks $T_{\BT}$ 
of the form in~\eqref{block-triangular-map-simplified} where $T_{\Yc}$ is the identity 
by choosing the reference to be $\source_\W = \target_\Y \otimes \source_\U$. Moreover, it is often desirable to choose $\source_\U = \target_\X$ so that the map 
$T_\Xc(y, \cdot)$ pushes the prior $\target_\X$ to the posterior distribution $\target_{\X \mid y}$ as in~\eqref{eq:triangular-conditioning-simplified}. Given the learned map, we generate posterior samples in parallel by evaluating the map at reference samples:
\begin{equation} \label{eq:ConditionalSamplingMaps}
    T_{\Xc}(\y,\x^{(i)}) \sim \target_{\X|\y}, \qquad \x^{(i)} \sim \target_{\X}.
\end{equation}
When the prior density $\target_\X$ is available in closed form and the transport map is invertible (see Section~\ref{sec:parameterization} for map parameterizations), we can also evaluate the posterior density using the change-of-variables formula~\eqref{eq:ChangeOfVariables}: 
$$\target_{\X|\y}(\x) = \target_\X(T_{\Xc}^{-1}(\y, \x))|\nabla_\x T_\Xc^{-1}(\y, \x)|,$$ 
where $T_{\Xc}^{-1}(\y , \cdot)$ denotes the inverse of $\x \mapsto T_{\Xc}(\y, \x)$. 

\rev{Alternative simulation-based inference methods do not directly approximate the posterior. Neural likelihood estimators~\cite{papamakarios2019sequential} learn a model for the conditional distribution $\target_{\Y|\X}$, which itself can be parameterized by a (block-)triangular map, while likelihood-ratio estimators~\cite{hermans2020likelihood} learn the ratio $\target_{\Y|\X}/\target_\Y$ using a classifier. These approaches can be useful when the posterior is not easily represented by a triangular map, but they typically require an additional sampling step, e.g., MCMC, for each observation. A common challenge to all of these approaches is that observations $\y$ are often high-dimensional, and low-dimensional summary statistics must be chosen or learned~\cite{chen2021neural}; see~\cite{cranmer2020frontier} for a broader overview.}

The methodology based on block-triangular maps has been used to solve the general inverse problems and sequential inference problems that were introduced in Sections~\ref{sec:BIP} and~\ref{sec:sequential_inference}, respectively. 
Some application domains include structural health monitoring~\cite{grashorn2024efficiency}, micro-structure reconstruction~\cite{dureth2023conditional} in mechanics, block copolymer self-assembly in material science~\cite{baptista2024bayesian}, reconstruction of turbulent flows~\cite{li2023multi}. 
and sequential inference (i.e., filtering) of chaotic dynamical systems~\cite{spantini2022coupling, taghvaei2022optimal}. Similarly, conditional generative models fundamentally solve the same problem by learning maps that can sample structured data given a conditioning variable, such as images from a text prompt, given the same assumption of paired data samples for learning~\cite{rombach2021highresolution, mirza2014conditional}.
Finally, triangular transport is also useful in optimal experimental design~\cite{huan2024optimal} where the focus is to identify observations $\y$ that are maximally informative of $\x$ by maximizing the information gain from the prior to the posterior. 
One measure is the mutual information between $\X$ and $\Y$, which is computed from the conditional distributions $\target_{\X|\y}$ for different realizations of $\y$. Triangular maps are natural for estimating mutual information since they do not require a separate sampling procedure for each value of $\y$ after training; see~\cite{li2024expected, koval2024tractable} for more details.

\subsection{Composed Maps and Residual Resampling} \label{sec:composed_maps}

We now describe a procedure to improve the quality of conditional samples from an imperfect (block)-triangular map. This is useful in settings where an estimated map $\widehat{T}$ only approximately pushes forward $\source_\W$ to $\target_\Z$. As a result, evaluating $\hatT_{\Xc}(\y,\cdot)$ at reference samples from $\source_\U$ as in~\eqref{eq:ConditionalSamplingMaps} only results in approximate samples from $\target_{\X|\y}$. Here we propose to use reference samples from 
the distribution $\widehat{\source}_{\U,\V} \coloneqq \hatT^{-1}\# \target_{\Y,\X}$ implied by pushing forward target samples through the inverse map. That is:
\begin{equation} \label{eq:ApproximateRefSamples}
(v^{(i)},u^{(i)}) = \hatT^{-1}(\y^{(i)},\x^{(i)}) \sim \widehat{\source}_{\V,\U}
\end{equation}
for $(\y^{(i)},\x^{(i)}) \sim \target_{\Y,\X}$. We then generate approximate conditional samples from $\target_{\X|\y}$ using $\hatT_{\Xc}(\y,\u^{(i)})$. \rev{Here and below we assume that $T_\Yc = \mathrm{Id}$ as in~\eqref{block-triangular-map-simplified}, so that $\u^{(i)} = \hatT_\Xc^{-1}(\y^{(i)},\x^{(i)})$.} 

By combining~\eqref{eq:ConditionalSamplingMaps} and~\eqref{eq:ApproximateRefSamples} we now define a \emph{composed map} $\hatT_c(\cdot,\cdot;y^\ast)\colon \Yc \times \Xc \rightarrow \Xc$ pushing forward the joint target distribution $\target_{\Y,\X}$ to the conditional distribution $\target_{\X \mid \y^\ast}$ for any $\y^\ast$ of the form:
\begin{equation} \label{eq:stochastic_map}
\hatT_{c}(\y,\x;\y^\ast) \coloneqq \hatT_{\Xc}(\y^\ast,\cdot) \circ \hatT_{\Xc}^{-1}(\y,\x),
\end{equation}
The composed map 
is also referred to as a \emph{stochastic map} in~\cite{spantini2022coupling} due to the map's dependence on the random variable $\Y$. 
\cite[Chapter~5]{baptista2022probabilistic} compares the pushforward distribution of $\target_{\Y,\X}$ through $\hatT_c$ 
to the pushforward distribution of the ideal reference $\source_\U$ through $\hatT_\Xc$ alone. In particular for Gaussian target distributions $\target_{\Z}$, the analysis proves that samples computed from~\eqref{eq:stochastic_map} using $N$ joint samples produce an estimator for the posterior covariance that converges at the faster $\mathcal{O}(1/N)$ rate, rather than the  $\mathcal{O}(1/\sqrt{N})$ rate when using $\hatT_\Xc$ alone. A related approach proposed in~\cite{tabak2020conditional} maps samples from all conditional distributions, i.e., $x^{(i)} \sim \target_{\X|\y^{(i)}}$ for all $\y^{(i)}$ to the Wasserstein barycenter of the conditional distributions before pushing the samples to the conditional of interest $\target_{\X|\y^\ast}$. 

Sampling from the reference space implied by the learned model, rather than the one assumed when learning the map is an example of \emph{residual resampling} or \emph{bootstrapping}~\cite{efron1994introduction}. The following remark clarifies this connection for a map of the form in Remark~\ref{rem:supervised_learning}. 

\begin{remark}[Connection to Residual Resampling] 
Given $N$ i.i.d.\ joint samples $(\y^{(i)},\x^{(i)}) \sim \target_{\Y,\X}$, we can fit the model $\x^{(i)} = \widehat{f}(\y^{(i)}) + \eta^{(i)}$ for some $\widehat{f}\colon\Yc \rightarrow \Xc$ using least-squares regression, assuming the residuals $\eta^{(i)}$ are independent and follow a mean-zero Gaussian distribution with variance $\sigma^2$. This yields the Gaussian approximation $\widehat\target_{\X|\y} = \mathcal{N}(\widehat{f}(\y),\sigma^2 I)$ to the conditional distribution.

To account for the true residuals being non-Gaussian, regression adjustment, popularized in approximate Bayesian computation~\cite{sisson2018handbook}, instead computes the empirical residuals $\eta^{(i)} := \x^{(i)} - \widehat{f}(\y^{(i)})$ 
and generates the samples
\begin{equation} \label{eq:ComposedRegression}
\widehat{f}(\y) + \eta^{(i)} = \widehat{f}(\y) + \x^{(i)} - \widehat{f}(\y^{(i)}),
\end{equation}
for $i=1,\dots,N$. These samples correspond to evaluations of the composed map $T_{c}$ when $T_\Xc$ has the form in Remark~\ref{rem:supervised_learning} and $\widehat{f}(y)$ is an estimate of the conditional mean $\mathbb{E}[\X|\Y = y]$. The composed map in~\eqref{eq:stochastic_map} generalizes~\eqref{eq:ComposedRegression} to sample from the full conditional distribution.
\end{remark}


Composed transport maps in~\eqref{eq:stochastic_map} form the basis of several ensemble filtering methods (see Section~\ref{sec:intro}) that transform samples from the prior distribution 
at each assimilation time to the filtering distribution. 
When $T_c$ takes the form in~\eqref{eq:ComposedRegression}, the maps correspond to the conditional mean filter in~\cite{Truong-Vinh2023}. \cite{spantini2022coupling} also shows that the popular ensemble Kalman filter corresponds to the special case when $T_c$ is an affine transformation of both inputs $(\y,\x)$. 

All of these approaches typically seek the map first and then compose it. One example of an approach that estimates the map via its composition is the MMD-based approach in \cite{dance2024causalinferencecocycles}, which minimizes a loss (over $\hatT_c$ as in \eqref{eq:stochastic_map}) equivalent to 
\begin{align*}
    & \mathbb{E}_{((\y,\x),(\y',\x'),(\y'',\x''))\sim \target_{\Y,\X}^{(3)}}[ k(\hatT_c(\y,\x;\y''), \hatT_c(\y',\x';\y'')) \\
    & \qquad + k(\x'',\x'') - 2 k(\x'', \hatT_c(\y,\x;\y''))]\;,
\end{align*}
where $\target_{\Y,\X}^{(j)}$ is the $j$-fold product measure; 
see Section~\ref{sec:tmi:scm:cf} for discussion on the application of composed maps to causal inference problems. 
Further development and study of objectives that directly seek composed maps is an interesting avenue of future work. 

\section{Causal Modeling and Inference} \label{sec:causal_inference}


In this section, we review recent work on causal modeling that is 
derived from and expressed in the language of triangular transport maps. For ease of presentation, we use the language of structural causal models (SCMs) and 
directed acyclic graphs (DAGs)  \citep{Spirtes2000,Pearl_2009} but we will adopt our previous transport notation which is
non-standard in the causal modeling literature. This change of notation further highlights the deep connections between the field of causal inference and transport. In Section~\ref{sec:structural_causal_models} we use triangular maps to parameterize and estimate SCMs while Sections~\ref{sec:interventions} and \ref{sec:tmi:scm:cf} describe how conditional transport maps can be used to model interventions and counterfactual quantities, respectively.

\subsection{Structural Causal Models via Triangular Maps} \label{sec:structural_causal_models}

Transport-based methods offer a natural framework to express causal models, assumptions, and interventions. In particular, the one-dimensional KR transport~\eqref{eq:OneDimMap} constructed from the quantiles and distribution functions appears in a variety of settings in causal inference including structural distribution models \cite{Vansteelandt_2014}, non-linear difference-in-differences \cite{athey_imbens_2006_DiD}, and quantile treatment effect estimation \cite{Chernozhukov_Hansen_2005_IV_quantile,chernozhukov2020instrumentalvariablequantileregression,sun2021qte}. 
In higher dimensions, 
triangular maps connect closely with SCMs and DAGs; 
see \cite{causalMLbook2025} for  an accessible introduction to SCMs.

\begin{figure}[htp]
    \centering
     \begin{overpic}[width=.6\linewidth, clip=true, trim= 6cm 5cm 12cm 5cm]{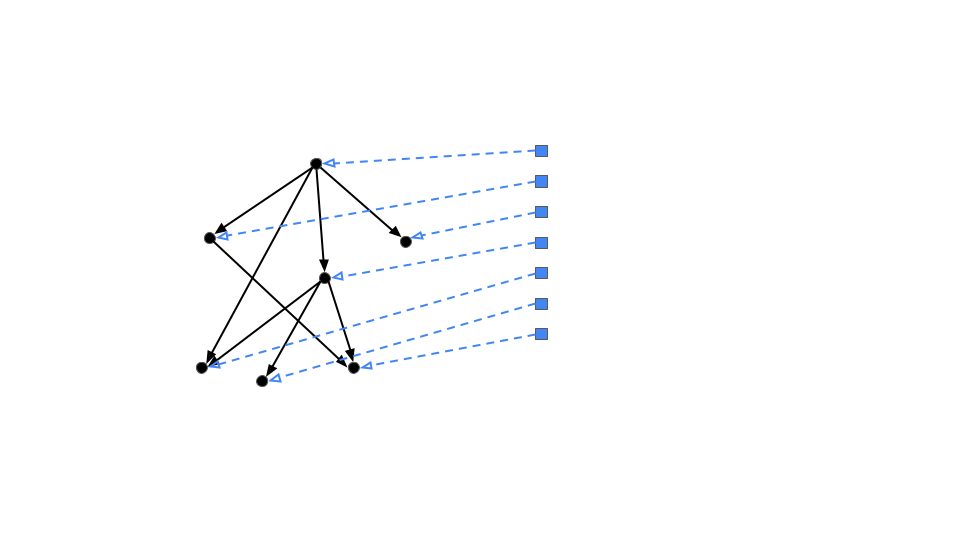}
        \put(30,55){$Z_1$}
        \put(3,39){$Z_2$}
        \put(52,38){$Z_3$}
        \put(26,28){$Z_4$}
        \put(38,1){$Z_5$}
        \put(18,-3){$Z_6$}
        \put(6,1){$Z_7$}

        \put(85,53){\scriptsize $W_1$}
        \put(85,46){\scriptsize $W_2$}
        \put(85,39){\scriptsize $W_3$}
        \put(85,33){\scriptsize $W_4$}
        \put(85,26){\scriptsize $W_5$}
        \put(85,19){\scriptsize $W_6$}
        \put(85,12){\scriptsize $W_7$}
    \end{overpic}
    \vspace{2ex}
    \caption{An example SCM represented with a DAG with $d=7$ 
    observable
    variables $Z_k$'s and the same number of noise variables $W_k$'s. Arrows pointing towards a 
    node indicate a causal relationship between the $Z_k$'s. The $W_k$'s 
    are independent noise variables injected in each node. In this 
    instance $Z_1$ is a parent of $Z_4$, i.e., $\pa_4 = \{ 1\}$ 
    while $Z_1, Z_4$ are parents of $Z_7$, i.e.,  $\pa_7 = \{1, 4\}$.}
    \label{fig:DAG}
\end{figure}

Consider a DAG that expresses the causal relationships between a set of variables $Z = (Z_1,Z_2,\dots,Z_d)$, where a directed edge from $Z_k$ to $Z_{k'}$ indicates that $Z_k$ is a direct cause of $Z_{k'}$ as shown in \Cref{fig:DAG}. The variables are indexed so as to give a valid topological ordering with respect to the DAG; that is, if there is a directed edge from $Z_k$ to $Z_{k'}$ then $k < k'$. Given this ordering, a SCM is specified by a set of stochastic 
relationships
\begin{align} \label{eq:sem}
    Z_k(z_{\pa_k}) := T_k(z_{\pa_k}, \Noise_k)\;, \quad k =1,\dotsc,d \;,
\end{align}
where $\pa_k$ is the index set of the parents of $Z_k$ in the DAG, i.e., those 
variables that have an edge terminating in $Z_k$. If $Z_k$ has no parents then $\pa_k = \emptyset$ and $Z_k = T_k(\Noise_k)$. Moreover, the ``noise'' variables $\Noise = (W_1,\dots,W_d)$ 
are assumed to be mutually independent, reflecting the unobserved influence of variables that are exogenous to the system. (This can be relaxed in some settings but we assume independence here.) 
Since any set of random variables whose joint distribution factorizes according to the conditional independencies encoded by the DAG can be expressed this way, two more assumptions are needed for causal modeling. The first is a consistency condition, namely, that the observed variables $(Z_k)_{k=1}^d$ are produced by generating the noise variables and passing them into the structural equations in \eqref{eq:sem}. The second is an ``invariance'' condition: each structural function $T_k$ remains the same under interventions to the other variables. 

\vspace{1ex}


For simplicity, assume that $Z_k \in \R$ for each $k$. Note that the topological ordering requires that $Z_k$'s parents have indices $< k$. Hence it is straightforward to check that recursively rewriting each structural function in~\eqref{eq:sem} in terms of the noise variables yields an equivalent representation of the SCM as $Z = T(\Noise)$, 
where $T$ is a triangular map and
$T_k$ is a function of only the first $k$ noise variables as in~\eqref{KR-map-component}. If $Z_k \in \R^{d_k}$ with $d_k > 1$ then $T$ is block-triangular as
in Section~\ref{sec:block-KR} below. Furthermore, assuming that each structural function $\noise_k \mapsto T_k(z_{\pa_k},\noise_k)$ is a monotonic increasing function of $\noise_k$ makes $T$ a triangular and monotone map. 
While there is nothing inherently causal about monotonicity, it is a common assumption made for practical purposes such as identifiability or ease of model specification. For example, the widely used additive noise model in terms of a mean function $m_k$,
\begin{align*}
    T_k(z_{\pa_k},\noise_k) = m_k(z_{\pa_k}) + \noise_k \;,
\end{align*}
is monotonic in $\noise_k$, as are many other commonly used functional forms
in the literature \citep{peters2014causal,immer2023identifiability,xi2025distinguishingcauseeffectcausal}.



In light of our formulation above, it is natural to model an SCM 
from empirical data using a triangular map. Indeed 
such models have been implemented as DAG-structured triangular normalizing flows \citep{pawlowski2020deep,khemakhem2021causal,javaloy2023causal,chen2023structured} estimated via maximum likelihood; see also Section~\ref{sec:parameterization}. 
As mentioned above, monotonicity of $T$ leads to an identifiable SCM, up to component-wise bijections of the exogenous noise variables;  see \cite{xi2023indeterminacy,javaloy2023causal} for details. 

The statistical properties of this procedure are not well understood, particularly in light of its subsequent use in estimating causal effects (as described below). As pointed out in \cite{dance2024causalinferencecocycles}, if the noise distribution is poorly matched to the distribution of the observations, flow-based estimators can have poor statistical properties. Even flexible neural network models may be mis-specified due to support or tail mis-match \cite{jaini2020tails,liang2022fat}, or require highly complex transport maps to match the distributions, resulting in slow rates of convergence \cite{irons2022triangular}. 


\subsection{Interventional  Estimands} \label{sec:interventions} 

Suppose we are interested in estimating the expected effect of an intervention
within an SCM. For example, one may be interested in the effect on a particular health outcome of a treatment at dosage $\phi$, or the effect on personal savings of increasing income by an amount $\phi$. Abstractly, let us write  $Z^{\phi}$ 
to denote the variables after the intervention $\phi$, 
with  $\mu_Z^{\phi}$ denoting its law. Then we may be interested in a  causal quantity that can be written generically as $\gamma(\target_Z^{\phi},\target_Z)$. For example, the average treatment effect (ATE) on variable $Z_k$ can be written
\begin{align*}
    \gamma(\target_Z^{\phi},\target_Z) \coloneqq \mathbb{E}[Z_k^{\phi}] - \mathbb{E}[Z_k] \;.
\end{align*} 
In the SCM framework with fixed noise distribution, an estimated composite structural function ${\widehat{\bff}}$ can be used to estimate the ATE and other interventional distribution estimands by generating samples $Z_k^{\phi}$ from $\mu^{\phi}_{Z_k}$ using
\begin{align*}
    Z_k^{\phi} = \widehat{\bff}_k(Z_{\pa_k}^{\phi}, \Noise_k) \;.
\end{align*}
In practice, this may involve inverting ${\widehat{\bff}}$ to modify the relevant components of $\bfN$ \cite{khemakhem2021causal,javaloy2023causal}. 
Beyond the average, quantities such as the interventional density may be estimated. In some cases, the samples themselves may be of interest. 

\paragraph*{Debiased Estimation} The plug-in estimators described above are versatile, but bias or slow rates of convergence in estimating $\hatT$ can be inherited by the estimator of interest. To address this, the well-developed methods for debiased estimation of scalar estimands such as the ATE \cite{kennedy2022semiparametric} have been extended to interventional density estimation \cite{Kennedy2023semiparametric,melnychuk2023normalizing,martinez2024counterfactual}, and more recently to (conditional) transport-based generative models for interventional distributions \cite{luedtke2025doublegen,dance2026debiasedcounterfactualgenerationflow}. These debiasing techniques allow state-of-the-art methods such as diffusion models and flow matching to generate confounding-corrected interventional samples of high-dimensional outcomes like text or images. Moreover, it was shown in \cite{dance2026debiasedcounterfactualgenerationflow} that the choice of reference distribution can impact the performance of such methods. In particular, the observational conditional distribution can be similar to the interventional distribution, especially under weak confounding, resulting in low-complexity transport maps between them. Using the observational distribution as a flow base simplifies the estimated transport maps, but requires an extension of the usual debiasing methods, including estimation of additional quantities. A full statistical characterization of the trade-offs remains an open problem.


\subsection{Counterfactual Estimands} \label{sec:tmi:scm:cf}

Unlike interventional quantities, counterfactual quantities require so-called ``cross-world'' assumptions. Although they have been criticized as unfalsifiable and therefore unscientific \cite{dawid2000}, nevertheless they are required in many settings of interest, from personalized medicine to economic and health policy. While we do not intend to take a philosophical stance on the topic here, we will discuss how conditional transports can be used to identify and estimate model-based counterfactual estimands if one is willing to defend the necessary assumptions. 

Estimating counterfactual quantities with SCMs is conceptually straightforward 
within a transport model. For example, Pearl's \emph{abduct, act, predict} procedure \citep{Pearl_2009}
can be stated as follows: 
\begin{itemize}
    \item (\emph{abduct}) Infer the distribution of the noise variables conditioned on the factual observation $\bfx$, $\widehat \rho_{\bfN|\bfX=\bfx}$;
    \item (\emph{act}) Perform the desired intervention on the factual, $\bfx \rightsquigarrow \bfx^{\phi}$;
    \item (\emph{predict}) Compute the desired counterfactual variables $Z^{\phi}_k = T_k(z^{\phi}_{\pa_k}, \widehat \Noise_k)$, where $\widehat \Noise_k \sim \widehat \rho_{\bfN|\bfX=\bfx}$.
\end{itemize}

If the SCM equations in \eqref{eq:sem} are bijective functions of $\Noise_k$, then for each $z_{\pa_k}$, the structural functions can be inverted in their second argument, 
allowing the abduction step to be performed in closed-form, without sampling, as 
\begin{align} \label{eq:bcm:abduct}
    \widehat{\Noise}_k = T_k^{-1}(Z_{\pa_k}, Z_k) \;,
\end{align}
where the inverse applies only to the second argument of $T_k$. This can be thought of as obtaining a generalized residual, in the sense that for linear additive noise models, $\widehat{\Noise}_k$ is the residual of regressing $Z_k$ on $Z_{\pa_k}$ akin  to Section~\ref{sec:composed_maps}. 
This  allows the intervened counterfactuals to be identified using the composed map 
\begin{align} \label{eq:cf:coupling}
    Z^{\phi}_k = T_k(z^{\phi}_{\pa_k}, \widehat{\Noise}_k) = T_k(z^{\phi}_{\pa_k}, T_k^{-1}(Z_{\pa_k}, Z_k))  \;.
\end{align}
Observe that the composed map is coherent across different intervention levels, in the sense that for any $z_{\pa_k}^{\phi},z_{\pa_k}^{\phi'}$,
\begin{align*}
    & Z_k^{\phi'}  = T_k(z_{\pa_k}^{\phi'}, T_k^{-1}(z_{pa_k}^{\phi},Z_k^{\phi})) \\ 
    & = T_k(z_{\pa_k}^{\phi'}, T_k^{-1}(z_{pa_k}^{\phi},\cdot))
    \circ T_k(z^{\phi}_{\pa_k}, T_k^{-1}(Z_{\pa_k}, Z_k))) \nonumber \\
    & = T_k(z_{\pa_k}^{\phi'}, T_k^{-1}(Z_{\pa_k}, Z_k)) \;. \nonumber
\end{align*}
It was shown in \cite{dance2024causalinferencecocycles} that the only counterfactual transport maps that are coherent across interventions are composed maps as in \eqref{eq:cf:coupling}, putting transport-based models of counterfactuals in correspondence with bijective SCMs. 

Naturally, the expressions above can also be estimated by replacing $T_k$
with $\widehat T_k$, which is estimated from empirical data (assuming that there are no unobserved confounders).
This procedure is used to estimate quantities that depend on the joint distribution between counterfactuals~\cite{khemakhem2021causal,nasr2023counterfactual}, including 
fairness-based counterfactuals \cite{javaloy2023causal,delara2024transport,Machado:2024aa}. 
Alternatively, the composed map (as in Section~\ref{sec:composed_maps}) 
\begin{align*}
    {\bff}_k(z_{\pa_k}^{\phi},\cdot)\circ \bff_k^{-1}(z_{\pa_k},z_k)
\end{align*}
can be estimated directly. Doing so targets the counterfactual transport without an intermediate reference distribution, eliminating the need to specify a reference distribution for the exogenous noise variables. It has been shown that poorly matched reference and target distributions can result in model mis-specification or require high-complexity transport maps that exhibit slow rates of convergence \cite{jaini2020tails,laszkiewicz2022marginal,liang2022fat,irons2022triangular}. Hence, targeting the composed map can mitigate those problems. Moreover, because the composed map quotients out the intermediate reference distribution, it is well-specified using lower complexity function classes, which can result in improved statistical performance, as discussed in detail in \cite{dance2024causalinferencecocycles}. A better theoretical understanding of this improvement is an open avenue of research. 



When directly targeting the coupling between counterfactual distributions, it has been argued that OT maps are natural, particularly in light of the uniqueness of the Brenier map (see \cref{thm:Brenier-theorem}). This is the approach of \cite{Black_2020,delara2024transport}, who also appeal to the philosophical argument \cite{lewis1973causation} that counterfactual counterparts should be as similar as possible while exhibiting the required change. When $Z_k \in \R$, OT is equivalent to using the monotone coupling. However, because the composition of two \emph{multivariate} OT maps generally does not yield another OT map \citep{santambrogio2015optimal}, multivariate OT couplings do not compose in a coherent way across more than two counterfactual settings (e.g., more than two treatment levels). This invalidates their use as counterfactual models in multi-level, multivariate settings. A related approach is to use OT maps between counterfactual distributions and a shared reference distribution \cite{ribeiro2025counterfactual}. This solves the coherence problem, but these maps are identified only up to the choice of reference distribution. This issue is discussed in more detail in \cite{dance2024causalinferencecocycles,Lara:2025aa:matching}. We note that these limitations for modeling counterfactuals do not preclude the use of OT maps for estimating interventional quantities \cite{TorousGunsiliusRigollet_2024} or bounding certain counterfactual estimands \cite{balakrishnan2025conservative,lin2025tightening,lin2026causal}.

\section{Methodological Extensions of Block-Triangular 
Maps} \label{sec:extended_methodology}

In this section we present two novel methodological extensions of triangular maps towards future research in sampling algorithms. Section~\ref{sec:block-KR} presents an extension of the KR map to act on blocks of variables in order to exploit hierarchical structure in groups of variables and to extend the 
map construction to abstract vector-valued random variables.  Section~\ref{sec:conditioning-dist-on-dist} describes how the methodology of triangular 
transport can be applied in more abstract settings to build maps that condition on distributional variables.



\subsection{The Block-KR Map}\label{sec:block-KR}

Here we outline a block extension of the KR map by combining our construction of block-triangular maps in~\Cref{sec:block-maps} with that of the KR maps in~\Cref{sec:KR_map}. Let us consider an abstract vector space $\Zc = \bigtimes_{i=1}^d \Zc_i$, as the product space of individual vector spaces $\Zc_i$ and consider a target distribution on $\Zc$. Then, the law of total probability and in turn the identity~\eqref{total-probability-R-d} still hold with the difference that now $Z = (Z_1, \dots, Z_d) \in \Zc$ is a ``block" vector with components $Z_i \in \Zc_i$ that are themselves vector valued random variables. Nevertheless, the main idea behind our construction of the KR map, i.e., identifying individual mappings between reference and target conditionals, remains valid in this case. 

For simplicity, suppose the reference has the tensor product form $\source_W = \bigotimes_{i=1}^d \source_{W_i}$, i.e., an independent product of the marginal distributions $\source_{W_i} \in \P(\Wc_i)$. By continuing our analogy with the construction of the KR maps, we can consider a block-KR map 
\begin{equation*}
    T_\BKR(w) = \Big( T_1(w_1), T_2(z_1, w_2), \dots, T_d(z_{1:d-1}, w_d) \Big) 
\end{equation*}
with components $T_k \colon (\times_{i=1}^{k-1} \Zc_i) \times \Wc_k \to \Zc_k$ that solve the individual transport problems 
\begin{equation*}
    T_k(z_{1:k-1}, \cdot) \# \source_{\W_k} = \target_{Z_k \mid z_{1:k-1}}.
\end{equation*}
One can take these maps to be Brenier maps exactly as we did in~\Cref{sec:regularity}, or any other \rev{OT} map.
Now, recursively define the maps 
$$
T_{1:k}(w_{1:k}) := \big( T_{1:k-1}(w_{1:k-1}), T_k( T_{1:k-1} (w_{1:k-1}), w_k\big)
$$
with the convention that $T_{1:1} = T_1$.
These maps are precisely block-triangular maps, in the parlance of~\Cref{sec:block-maps}, that push $\source_{W_k}$ forward to the conditional distributions of $\target_{Z_k \mid T_{1:k-1}(w_{1:k-1})}$.
In light of these observations, various algorithms can be proposed for the estimation of $T_{\BKR}$ maps using some of the ideas that we discussed in~\Cref{sec:computation} since the map components $T_k$ with vector-valued inputs  no longer have the simple form in~\eqref{eq:KRcomponent_parentdependence} for KR maps with scalar inputs.

\subsection{Conditioning Distributions on Distributions}
\label{sec:conditioning-dist-on-dist}
Recall that we introduced block-triangular maps in \Cref{sec:block-maps}
by thinking of the $T_\Xc(\y,\cdot)$ components as a parametric 
family of mappings between $\source_{\U}$ and $\target_{\X \mid \y}$. 
This parametric perspective allows us to take the elements $\y$ to be 
in very abstract spaces. In particular, we can take $\y$ to be a measure itself. 
More precisely, let $\Yc$ be the space of probability measures on a compact set $\Gamma \subset \R^n$  and let $\Xc$ be a finite dimensional space as before. Now we can consider a target measure $\target_{\Y,\X}$ where the marginal $\target_\Y$ is now the law of a random measure on $\Gamma$. We can now construct, as before, the reference $\source_{\V,\U} = \target_\Y \otimes \source_\U$ and consider block-triangular maps of the form \eqref{block-triangular-map-simplified}. \Cref{thm:triangular_conditional_pushforward} and the identity \eqref{eq:triangular-conditioning-simplified} remain valid, however, the map $T_\Xc(\cdot, y)$ is now a conditioning map or a generative model parameterized by a measure. This extension of triangular maps was utilized in the recent work~\cite{al2025fast}
to build an amortized filtering algorithm in which case the parameter $y$ was the empirical distribution of prior particles 
in a sequential inference problem; see~\Cref{sec:sequential_inference}. These problems have also appeared in the recent works~\cite{geshkovski2023mathematical,geshkovski2024measure} in mathematical models of the attention mechanisms in neural networks.
More broadly, the idea of learning models parameterized by distributions has existed in the machine learning literature for some time; see for example~\cite{muandet2012learning,oliva2014fast,szabo2016learning}.

\section{Outlook} \label{sec:outlook}

This article discussed triangular measure transport as a framework for \rev{conditional sampling}. While significant work has been dedicated to the field of computational \rev{OT}, triangular transport remains less explored while providing exciting new avenues for solving practical problems. We addressed two applications in this review for Bayesian inference and causal modeling. The key idea in triangular transport is that the transport map is defined explicitly from the conditional distributions of the reference and target distributions. First, the map thus inherits properties such as regularity and conditional independencies. Second and most importantly, the explicit construction from the marginal conditionals provides a mechanism to efficiently sample conditionals of the target distribution. Building on the foundational theory and methodologies of triangular transport presented in this review, there are many directions of future research. We describe some of these directions below.

\paragraph*{Theory of Triangular Transport}
While some of the ideas we discussed, such as the KR map, are classical, our broader theoretical understanding of triangular transport remains limited beyond 
our discussion in~\Cref{sec:transport_map}. In particular, understanding the regularity of KR maps (Section~\ref{sec:KR_map}) and block-KR maps (Section~\ref{sec:block-KR}) in relation to properties of the reference and target distributions remains open. This problem is of fundamental interest in the analysis of amortized inference algorithms based on triangular transport and in particular the analysis of errors and convergence properties of said algorithms. \rev{Moreover, the extensions of Theorems~\ref{thm:Brenier-theorem} and~\ref{thm:caffarelli-regularity} on the construction and regularity of Brenier maps on infinite-dimensional spaces are also of interest, particularly for function-space inference problems.}

\rev{The statistical results in Section~\ref{sec:error-analysis} also leave open several theoretical questions that have practical impact on the map's estimation error. These include how to select the coordinates or block-orderings of variables, the construction of estimators that adapt to sparsity and intrinsic dimension of the target distribution, two-sample estimation when both the reference and target distribution are prescribed using samples, and guarantees that combine approximation, sampling and optimization errors.}

\paragraph*{Methodologies for Triangular Transport}
While many algorithms exist for estimating triangular transport maps (Section~\ref{sec:computation}), several directions and extensions remain underexplored. One particular challenge with map estimation, going beyond triangular maps, is the issue of tails. Often, very few samples are available in the tails of the distributions. As a result, the estimated maps may be inaccurate and have unstable behavior in these regions. 
For example, it was shown in~\cite{jaini2020tails, liang2022fat} that normalizing flows using bi-Lipschitz maps are not able to capture target distributions if the tails of the reference are lighter \rev{or heavier} than the target. This question also 
relates to the choice of the reference on the computed transport map, which is largely open both theoretically and algorithmically. Work in causal inference \cite{dance2024causalinferencecocycles,dance2026debiasedcounterfactualgenerationflow} indicates that using an empirical distribution as the reference can mitigate this issue, but that may not always be possible, and a full theoretical understanding is still needed.



 Another interesting idea is to design maps that inherit certain structure 
 from the reference and target distributions. 
 For example, we may want KR maps whose components depend on a 
 sparse set of variables. In~\cite{spantini2018inference}, the conditional independence properties (i.e., Markov properties) of the target distribution or its marginal conditionals were shown to directly translate into sparse dependence of the transport map. This was used in~\cite{baptista2024learning} to learn the probabilistic graphical models from data by exploiting sparse transport representations for the target distribution. 
 More broadly, we can seek other forms of structure. These include identifying low-dimensional subspaces on which the target differs from the reference~\cite{brennan2020lazy}, so that the transport map need only be learned on those subspaces \rev{(i.e., a \emph{spiked transport} map~\cite{baptista2025dimension}),} or finding the variable orderings and groupings for hierarchical transports (of the form in Section~\ref{sec:block-KR}) that yield the greatest sparsity or are easiest to approximate.
 Incorporating physical constraints into transport maps for scientific computing is also an important direction for connecting transport to the field of physics-informed generative modeling~\cite{yang2020physics,yang2019enforcing,cheng2024hard}.

\paragraph*{Combining Transport With Sampling} \label{sec:transport_accelerated_MCMC}

The estimation of triangular maps naturally leads to errors due to 
limited data and algorithmic limitations. It is natural to ask whether 
asymptotically exact methods such as MCMC or importance sampling can be used to correct these errors. Conversely, we may ask whether it is possible to use triangular transport to 
improve the performance of other sampling algorithms. 

Given a target distribution $\mu_\Z$, one may first learn a transport map $T$ 
such that $T \# \source_\W \approx \target_\Z$ where $\source_\W$ is assumed to be ``nice", e.g., log-concave and that can be simulated efficiently using MCMC. Then, one can define a new target by pulling $\mu_\Z$ back to $\source_\W$, i.e., sampling from $T^{-1}\# \target_\Z$. If $T^{-1} \#\target_\Z$ is closer to $\source_\W$, then our MCMC algorithm can simulate this distribution efficiently with better mixing. We note that 
if the target density is known up to a normalizing constant, the pullback density can also be evaluated up to the same constant above and so this approach can work well for Bayesian inference tasks. This idea was explored in~\cite{parno2018transport, gabrie2022adaptive} to accelerate Metropolis-Hastings algorithms.
In practice, the map can be estimated simultaneously to running MCMC 
by using a subset of MCMC samples to learn and refine the map.

\paragraph*{Bayesian Inference}
While triangular transport for Bayesian inference is now well-established,
and has been applied in various practical problems \cite{baptista2024bayesian, grashorn2024efficiency, dureth2023conditional, li2023multi, spantini2022coupling} this framework has yet to be adopted for 
the solution of large-scale inverse problems in the broader scientific 
community. This adoption requires significant effort towards the methodological and computational aspects of transport models towards the scaling of this methodology 
to large and multi-modal data sets and complex PDE models. \rev{In the same vein, 
it is crucial to develop a clear understanding of how transport approaches 
compare to more traditional sampling and inference methods such as MCMC 
and variational inference. To the best of our knowledge, such a systematic numerical comparison on canonical problems does not exist in the literature.}

\paragraph*{Discovering Causal Relationships and Structure}

Our discussion regarding SCMs in \Cref{sec:causal_inference}
 assumed a known (or partially known) causal DAG was available. In many settings, 
 the causal graph is unknown and learning the graph itself from data is of interest. The earliest methods for so-called \emph{causal discovery} were based on conditional independence tests \cite{Spirtes2000}. Those ideas were combined with measure transport in \cite{akbari2023causal}. Recent works have shown that when the variables are in a valid topological order with respect to the underlying DAG, the non-zero entries of the Jacobian of the inverse of the KR map encode the edges of the DAG \cite{khemakhem2021causal,javaloy2023causal,xi2023triangular}. In practice, the order of the variables may also be unknown and must be searched over. By extending results from \cite{Raskutti2018,spantini2018inference}, it was shown in \cite{xi2023triangular} that the variable ordering (i.e., permutation of variable indices) with the sparsest corresponding Jacobian is equivalent to the true underlying DAG. However, efficiently searching over the variable ordering presents an important research challenge, as does inferring the zeros of the Jacobian. Developing tractable methods to identify ordering and structure will further improve the statistical performance and ultimate adoption of transport methods for conditional sampling problems. This also has implications for the counterfactual methods in Section~\ref{sec:tmi:scm:cf}: in practice TMI maps are used as coherent and identifiable counterfactual transports, which requires a known variable ordering. On the other hand, OT maps are equivariant with respect to index permutations, making them insensitive to variable ordering. To our knowledge, whether there exists a family of transport maps that is coherent, identifiable, and insensitive to variable ordering remains unknown. 

\section*{Acknowledgments}
RB is grateful for support from 
the Natural Sciences and Engineering Research Council of Canada (NSERC) under award RGPIN2026-07896. 
BBR gratefully acknowledges the support of NSERC under awards: RGPIN2020-04995, RGPAS-2020-00095.
BH was supported by the National Science Foundation grants
DMS-2208535, ``Machine Learning for Bayesian Inverse Problems'' and 
DMS-2337678, "CAREER: Gaussian Processes for Scientific Machine Learning: Theoretical Analysis and Computational Algorithms".

%% file: diagram.tex
\begin{figure}[!ht]
\scalebox{0.41}{\begin{tikzpicture}
  \node (J) {\includegraphics[width=12cm]{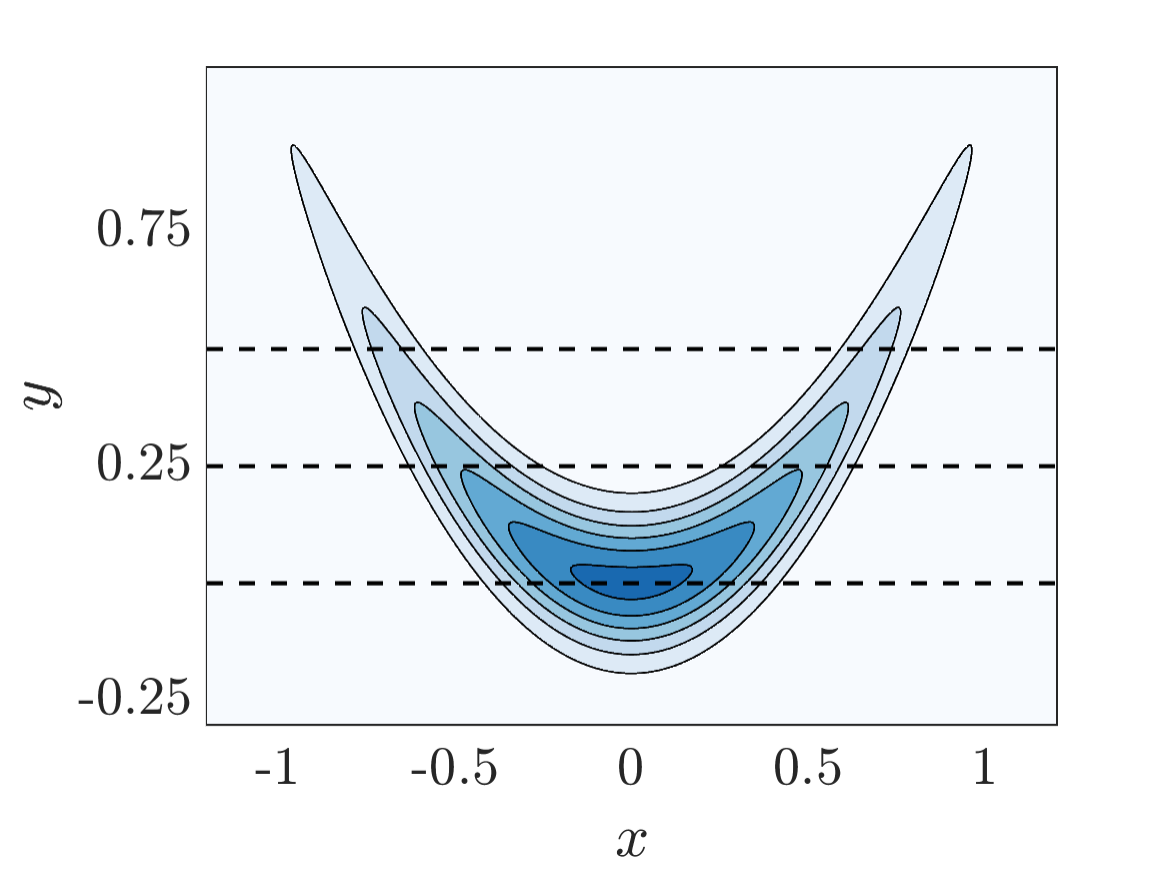}};

  \node[below=1mm of J] (P2) {
  \begin{overpic}[width=7cm]{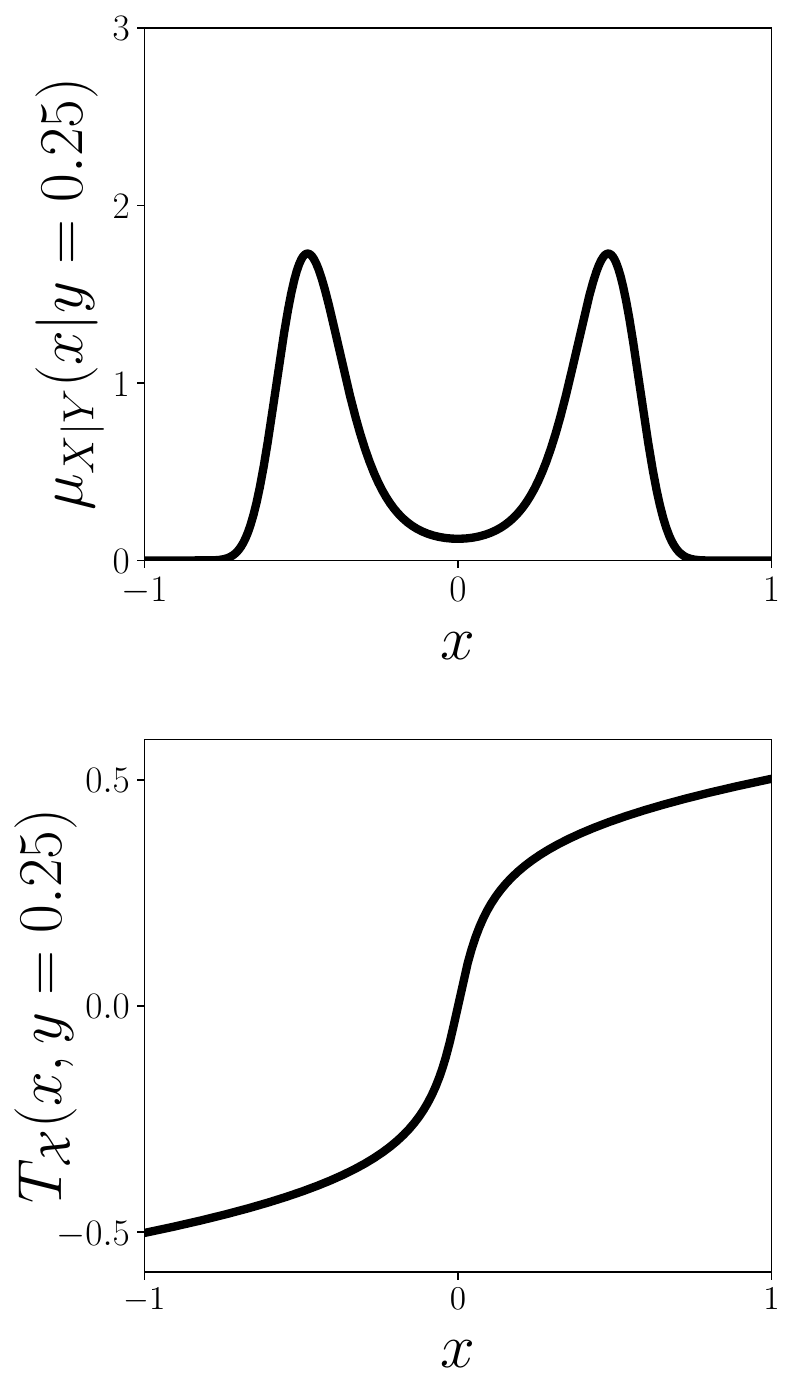}
     \put(32,85){\large B} 
  \end{overpic}
  };
  \node[right=1mm of P2] (P1) {
  \begin{overpic}[width=7cm]{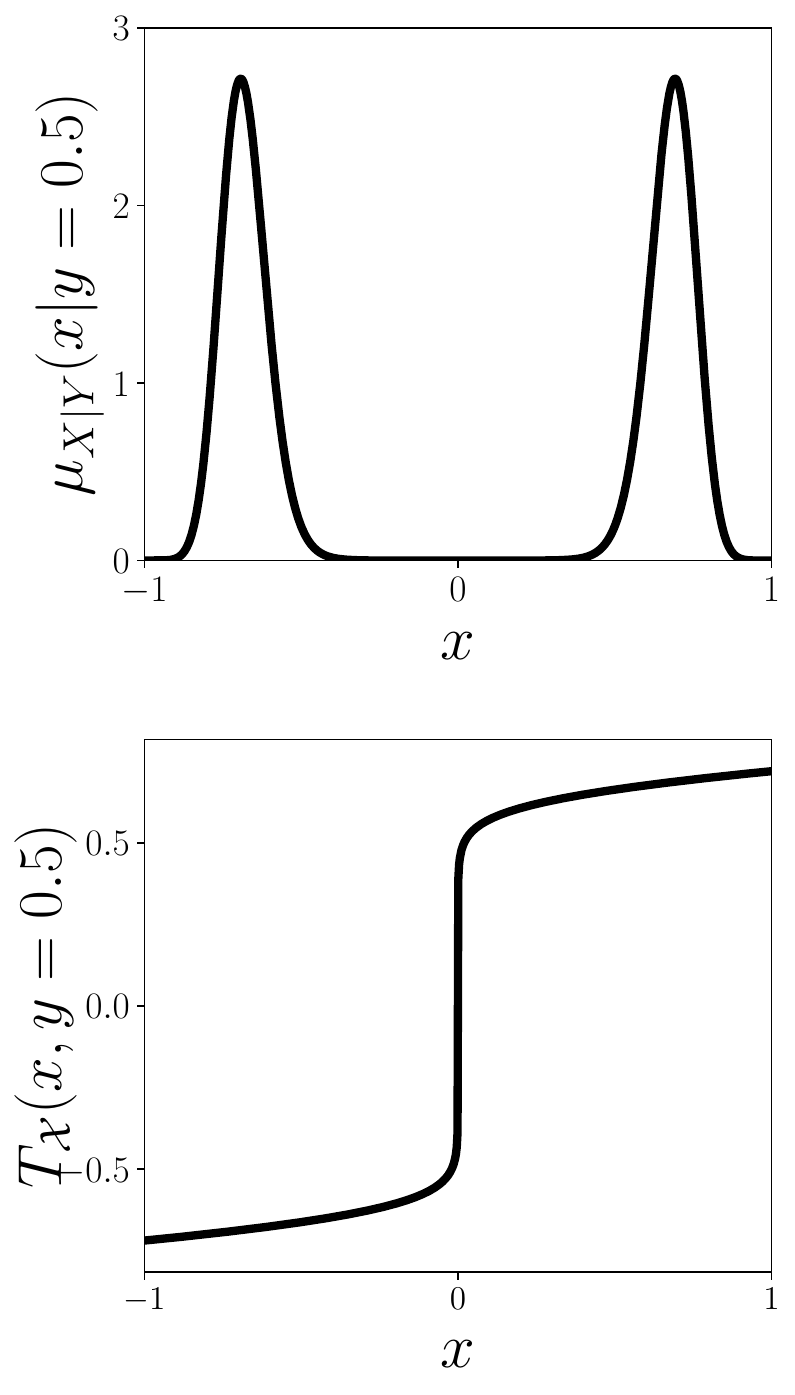}
     \put(32,85){\large C} 
  \end{overpic}
  };
  \node[left=1mm of P2] (P3) {
  \begin{overpic}[width=7cm]{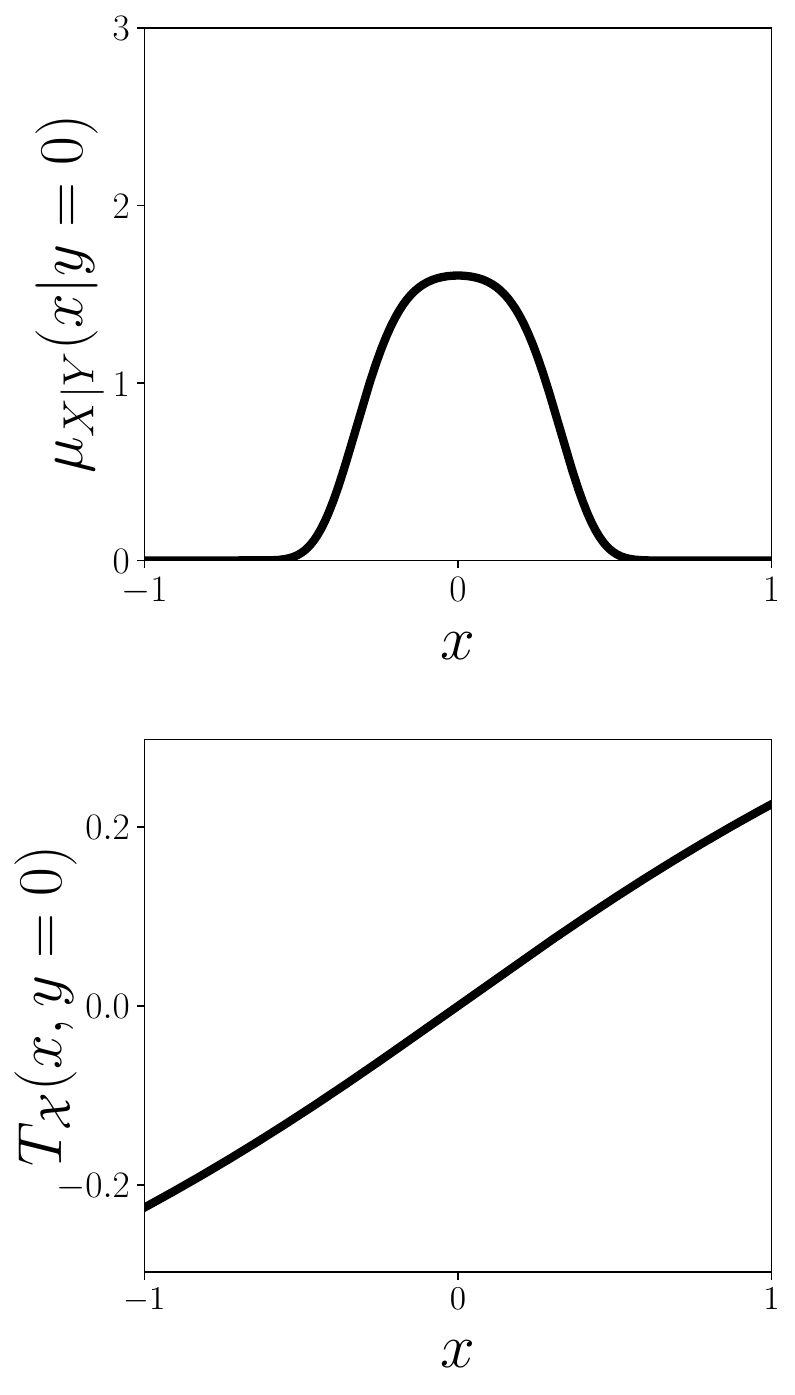}
     \put(32,85){\large A} 
  \end{overpic}
  };

  \node[left=0.4cm of J.east, yshift=-1.3cm] (C3) {\Large A};
  \node[left=0.4cm of J.east, yshift=-0.22cm] (C2) {\Large B};
  \node[left=0.4cm of J.east, yshift=0.8cm] (C1) {\Large C};

\end{tikzpicture}}
\caption{Joint distribution $\mu_{\X,\Y}$ for a quadratic likelihood model (Top), conditional distributions $\mu_{\X|\y}$ (Middle) and transport maps $T_\y$ (Bottom) for three realizations of the conditioning variable $\y$. \label{fig:example_cde}}
\end{figure}